\documentclass[12pt]{article}
\pdfoutput=1
\usepackage{amsmath}
\usepackage{epsfig}
\newcommand{\be}{\begin{equation}}
\newcommand{\ee}{\end{equation}}

\newcommand{\no}{\noindent}
\newcommand{\ce}{\begin{center}}
\newcommand{\nc}{\end{center}}

\makeatletter
\@addtoreset{equation}{section}
\makeatother
\renewcommand{\theequation}{\thesection.\arabic{equation}}

\def\sqr#1#2{{\vcenter{\vbox{\hrule height.#2pt
 \hbox{\vrule width.#2pt height#1pt \kern#1pt
 \vrule width.#2pt} \hrule height.#2pt}}}}

\def\operp{\hbox{${\kern+.25em{\bigcirc}
\kern-.85em\bot\kern+.85em\kern-.25em}$}}

\def\lsim{\;\raise0.3ex\hbox{$<$\kern-0.75em\raise-1.1ex\hbox{$\sim$}}\;}
\def\gsim{\;\raise0.3ex\hbox{$>$\kern-0.75em\raise-1.1ex\hbox{$\sim$}}\;}
\def\no{\noindent}

\def\ce{\centerline}
\def\ve{\vfill\eject}
\def\rdots{\mathinner{\mkern1mu\raise1pt\vbox{\kern7pt\hbox{.}}\mkern2mu
 \raise4pt\hbox{.}\mkern2mu\raise7pt\hbox{.}\mkern1mu}}

\usepackage{indentfirst}

\def\e e{$e^+ e^-$ }

\newcommand{\ket}[1]{\left| #1 \right>} 
\newcommand{\bra}[1]{\left< #1 \right|} 
 
\usepackage{slashbox}
\usepackage{pict2e}

\usepackage{slashed}
 
 \usepackage{mathrsfs}
 
\def\subequations{\refstepcounter{equation}%
\edef\@savedequation{\the\c@equation}%
\@stequation=\expandafter{\theequation}
\edef\@savedtheequation{\the\@stequation}
\edef\oldtheequation{\theequation}%
\setcounter{equation}{0}%
\def\theequation{\oldtheequation\alph{equation}}}%

\def\endsubequations{%
\setcounter{equation}{\@savedequation}%
\@stequation=\expandafter{\@savedtheequation}%
\edef\theequation{\the\@stequation}\global\@ignoretrue}

\DeclareMathAlphabet{\mathpzc}{OT1}{pzc}{m}{it}

\begin{document}

\ce{\bf An SLq(2) Extension of the Standard Model}
\vskip.3cm

\ce{\it Robert J. Finkelstein}
\vskip.3cm

\ce{Department of Physics and Astronomy}
\ce{University of California, Los Angeles, CA 90095-1547}

\vskip1.0cm

\no {\bf Abstract.}  We examine a quantum group extension of the standard model.  The field operators of the extended theory are obtained by replacing the field operators $\psi$ of the standard model by ${\psi}$D$^j_{mm'}$, where D$^{j}_{mm'}$ are elements of a representation of the quantum algebra SLq(2), which is also the knot algebra.  The D$^j_{mm'}$ lie in this algebra and carry the new degrees of freedom of the field quanta.  The D$^j_{mm'}$ are restricted jointly by empirical constraints and by a postulated correspondence with classical knots.  The elementary fermions are described by elements of the trefoil ($j=\frac{3}{2}$) representation and the weak vector bosons by elements of the ditrefoil ($j=3$) representation.  The adjoint ($j=1$) and fundamental ($j=\frac{1}{2}$) representations define hypothetical bosonic and fermionic preons. All particles described by higher representations may be regarded as composed of the fermionic preons.  This preon model unexpectedly agrees in important detail with the Harari-Shupe model. The new Lagrangian, which is invariant under gauge transformations of the SLq(2) algebra, fixes the relative masses of the elementary fermions within the same family.  It also introduces form factors that modify the electroweak couplings and provide a parametrization of the Cabbibo-Kobayashi-Maskawa matrix.  It is additionally postulated that the preons carry gluon charge and that the fermions, which are three preon systems, are in agreement with the color assignments of the standard model.

\thispagestyle{empty}
\setcounter{page}{0}

\newpage

\no {\bf Introduction}
\vskip 0.5cm
\no  The possibility that the elementary particles are knots has been suggested by many authors, going back as far as Kelvin.$^{(1)}$  Among the different field theoretic attempts to construct classical knots, a model related to the Skyrme soliton has been described by Fadeev and Niemi.$^{(2)}$  There are also the familiar knots of magnetic field; and since these are macroscopic expressions of the electroweak field, it is natural to extrapolate from macroscopic to microscopic knots of this same field.  One expects that the conjectured microscopic knots would be quantized, and that they would be observed as solitonic in virtue of both their topological and quantum stability.  It is then natural to ask if the elementary particles might also be knotted.  If they are, one expects that the most elementary particles, namely the elementary fermions, are also the most elementary knots, namely the trefoils.  This possibility is suggested by the fact that there are 4 quantum trefoils and 4 classes of elementary fermions, and is supported by a unique one-to-one correspondence between the topological description of the four quantum trefoils and the quantum numbers of the four fermionic classes.  We have attempted to determine the minimal restrictions on a model of the elementary particles in the context of weak interactions if the quantum knot is described only by its symmetry algebra SLq(2) independent of its field theoretic origin.  The use of this symmetry algebra to define the quantum knot is similar to the use of the  symmetry algebra of the rotation group to define the quantum spin.\\

\newpage

\centerline{\Large\bf{I. Kinematics}}

\section{The Knot Algebra and the Quantum Knot}

\indent Let  $\left( \begin{array}{cc} a & b\\ c & d\\ \end{array} \right)$ be a two-dimensional representation of SLq(2), the knot algebra.

\noindent Then
\begin{equation}
\begin{array}{rcl} 
ab & = & qba \\ ac & = & qca
\end{array}
\hspace{1.0cm}
\begin{array}{rcl}
bd & = & qdb \\ cd & = & qdc
\end{array}
\hspace{1.0cm}
\begin{array}{rcl}
ad-qbc & = & 1 \\ da-q_1cb & = & 1
\end{array}
\hspace{1.0cm}
\begin{array}{rcl}
bc & = & cb \\ q_1 & \equiv & q^{-1} 
\end{array}
\end{equation}

\noindent where we take $q$ real.

Let D$^{j}_{mm'}$ be a $2j+1$ representation of SLq(2)$^{(3)(8)}$.

\noindent Then
\begin{equation}
\label{1.2}
\mbox{D}^j_{mm'}(a,b,c,d)=\sum_{\begin{array}{rllll} 0 & \le & s & \le &  n_+ \\ 0 & \le & t & \le & n_- \end{array}} \mbox{A}^j_{mm'}(q,s,t) \delta (s+t,n'_+) a^s b^{n_+-s} c^t d^{n_--t}
\end{equation}

\noindent where
\begin{displaymath}
\mbox{A}_{mm'}^j \left(q,s,t \right) = \left[ \frac{\langle n_+^\prime \rangle_1 ! \langle n_-^\prime \rangle _1 !}{\langle n_+ \rangle_1! \langle n_- \rangle_1 !}\right]^{\frac{1}{2}} \frac{ \langle n_+ \rangle_1 !}{\langle n_+ - s \rangle_1 ! \langle s \rangle_1 !} \frac{\langle n_- \rangle_1 !}{\langle n_- - t \rangle_1 ! \langle t \rangle_1 !}
\end{displaymath}
and where
\begin{displaymath}
\langle n \rangle_1 = \frac{q_1^n -1}{q_1 -1} \hskip0.5cm \mbox{and} \hskip0.5cm \begin{array}{rcl} n_\pm & = & j \pm m \\ n_\pm^\prime & = & j \pm m^\prime \end{array}
\end{displaymath}

\noindent The algebra (1.1) is invariant under the gauge transformations:

\begin{equation}
\mbox{U}_a(1): \begin{array}{rcl} a' & = & e^{i \varphi_a}a \\ d' & = & e^{-i \varphi_a}d \end{array} \hskip 2.0cm \mbox{U}_b(1): \begin{array}{rcl} b' & = & e^{i \varphi_b}b \\ c' & = & e^{-i \varphi_b}c \end{array}
\end{equation}

\noindent Then U$_a$(1) $\times$ U$_b$(1) induces on D$^j_{mp}$(a,b,c,d) the gauge transformation$^{(3)}$
\begin{equation}
\mbox{D}^j_{mp} (a',b',c',d') = e^{i( \varphi_a + \varphi_b)m} e^{i(\varphi_a-\varphi_b)p}\mbox{D}^j_{mp}(a,b,c,d)
\end{equation}

\noindent or
\begin{equation}
\mbox{D}^{j \hskip 0.1cm '}_{mp} = \mbox{U}_m\times\mbox{U}_{p} \mbox{D}^j_{mp}
\end{equation}

\noindent If $\ket{n}$ is a ket lying in the state space of SLq(2), we define a quantum knot by the state function
\begin{equation}
\psi \mbox{D}^j_{mm'} \ket{n}
\end{equation}

\noindent where $\psi$ is a standard quantum mechanical state function and
\begin{equation}
(j, m, m') = \frac{1}{2}(N,w,r+o)
\end{equation}

\noindent Here ($N,w,r$) are the number of crossings, the writhe, and the rotation of the 2d-projection of the corresponding oriented 3d-classical knot.  The factor $\frac{1}{2}$ allows half integer representations of SLq(2).  Since 2$m$ and $2m'$ are of the same parity while $w$ and $r$ are topologically constrained to be of opposite parity, $o$ is an odd integer which we set $=$1 for a trefoil knot.$^{(3)}$

The knot degrees of freedom are confined to the D$^j_{mm'}$ factor and are introduced here similarly to the way spin degrees of freedom are introduced by adjoining a spin factor to a state without spin.  This definition of the quantum knot allows only those selected states ($j,m,m'$) of the full 2$j+1$-dimensional representation that are permitted by the ($N,w,r$) spectrum of the 2d-projection of the corresponding classical knot.  Eqn. (1.7) is the ``correspondence principle" of the model where $\left( j,m,m' \right)$ describes the quantum knot and $\left( N, w, r \right)$ refers to the corresponding classical knot.\\

\noindent Note: There are only two classical trefoils because classical states of opposite $r$ are not topologically distinguished; but states of quantum trefoils, as here defined, with opposite $r$ can be distinguished by different $m'$ (and are assigned different hypercharge in the present model.)

\section{The Knotted Standard Model}

To go from the standard model to the knotted standard model we try to replace every field operator, $\Psi$, of the standard model by

\begin{equation}
\hat{\Psi}^j_{mm'} = \Psi \mbox{D}^j_{mm'} (\Psi)
\end{equation}

\noindent where $\hat{\Psi}$ is the field operator of the knotted model.  Since $\mbox{D}^j_{mm'}$ lies in the SLq(2) algebra, (2.1) adds new degrees of freedom to the field quanta.

Under U$_a$ $\times$ U$_b$ transformations of the algebra (1.1) the new field operators transform, by (1.4), as follows:
\begin{eqnarray}
\hat{\Psi}^{j\hskip0.1cm '}_{mm'} &=& \Psi \mbox{D}^{j \hskip 0.1cm '}_{mm'} \\
\hskip1.3cm &=& \mbox{U}_m \times \mbox{U}_{m'} \hat{\Psi}^j_{mm'}
\end{eqnarray}

\noindent For physical consistency the new field action must be invariant under (2.3), since (2.3) can be induced by U$_a$ $\times$ U$_b$ transformations that leave the defining algebra unchanged.  There are then Noether charges associated with U$_m$ and U$_{m'}$ that may be described as writhe and rotation charges, Q$_w$ and Q$_r$, since $m=\frac{w}{2}$ and $m'$=$\frac{1}{2}(r+1)$ for trefoils.

We define
\begin{eqnarray}
Q_w &\equiv& -k_wm \hskip0.2cm (\equiv -k_w\frac{w}{2}) \\
\mbox{Q}_{r} &\equiv& -k_{r}m' \hskip 0.2cm (\equiv -k_{r}\frac{1}{2}(r+1))
\end{eqnarray}

\noindent where k$_w$ and k$_{r}$ are undetermined constants with dimensions of electric charge.

We now compare the so defined $Q_w$ and $Q_r$ charges of the four quantum trefoils with the charge and hypercharge of the four fermion families in Table 2.1. $^{(3)}$

\begin{center}
{\bf{\underline{Table 2.1}}}
\begin{tabular}[h]{ccccc|ccccc}
\multicolumn{5}{c}{\underline{Standard Model}} &
\multicolumn{5}{c}{\underline{Quantum Trefoil Model}} \\
\underline{$(f_1,f_2,f_3)$} & \underline{$t$} &
\underline{$t_3$} &
\underline{$t_0$} & \underline{$Q_e$} & \underline{$(w,r)$} &
\underline{$\mbox{D}^{N/2}_{\frac{w}{2}\frac{r+1}{2}}$} &
\underline{$Q_w$} & \underline{$Q_r$} & \underline{$Q_w+Q_r$} \\
$(e,\mu,\tau)_L$ & $\frac{1}{2}$ & $-\frac{1}{2}$ & $-\frac{1}{2}$
& $-e$ & (3,2) & $\mbox{D}^{3/2}_{\frac{3}{2}\frac{3}{2}}$ &
$-k\left(\frac{3}{2}\right)$ & $-k\left(\frac{3}{2}\right)$ &
$-3k$ \\
$(\nu_e,\nu_\mu,\nu_\tau)_L$ & $\frac{1}{2}$ & $\frac{1}{2}$ &
$-\frac{1}{2}$ & 0 & (-3,2) & $\mbox{D}^{3/2}_{-\frac{3}{2}
\frac{3}{2}}$ & $-k\left(-\frac{3}{2}\right)$ &
$-k\left(\frac{3}{2}\right)$ & 0 \\
$(d,s,b)_L$ & $\frac{1}{2}$ & $-\frac{1}{2}$ & $\frac{1}{6}$ &
$-\frac{1}{3}e$ & (3,-2) & $\mbox{D}^{3/2}_{\frac{3}{2}-\frac
{1}{2}}$ & $-k\left(\frac{3}{2}\right)$ & $-k\left(-\frac{1}{2}
\right)$ & $-k$ \\
$(u,c,t)_L$ & $\frac{1}{2}$ & $\frac{1}{2}$ & $\frac{1}{6}$ &
$\frac{2}{3}e$ & (-3,-2) & $\mbox{D}^{3/2}_{-\frac{3}{2}
-\frac{1}{2}}$ & $-k\left(-\frac{3}{2}\right)$ &
$-k\left(-\frac{1}{2}\right)$ & $2k$ \\  & & & & & & & $Q_w=-k_w\frac{w}{2}$ &
$Q_r = -k_r\frac{r+1}{2}$& \\
\end{tabular}
\end{center}

Note that with the particular row to row correspondence, $(f_1, f_2, f_3) \leftrightarrow (w, r)$, in this table, \underline{and only for this correspondence}, is there proportionality between the $(t_3,t_0,Q_e)$ and the $(Q_w, Q_r, Q_w+Q_r)$ columns.

\noindent To construct and interpret this table we have postulated that $k$ is a universal constant in the following sense:
\begin{equation}
k_w = k_r = k
\end{equation}
\noindent and
\begin{equation}
k = \frac{e}{3}
\end{equation}
with the same value for all trefoils.

\noindent Then it follows from the table that the total electric charge ($Q_w + Q_r$) of each trefoil is the same as the electric charge $Q_e$ of the corresponding family of fermions:$^{(3)(4)}$
\begin{equation}
Q_w + Q_r = Q_e
\end{equation}
and that
\begin{equation}
Q_w = et_3
\end{equation}
\noindent and
\begin{equation}
Q_r = et_0
\end{equation}
Then $\left(2.8\right)-\left(2.10\right)$ are  in agreement with the standard model for which
\begin{equation}
Q_e=e(t_3+t_0)
\end{equation}

\noindent Since these relations hold for the special ($f_1$, $f_2$, $f_3$) $\leftrightarrow$ $(N, w, r)$ row to row correspondence in the Table 2.1, and only for this particular match between the trefoils and the fermionic families, the correspondence itself, in addition to the value of $k$ as $\frac{e}{3}$, is empirically fixed and unique.

We accordingly identify the writhe charge, $Q_w$, of the trefoil with the isotopic charge of the standard fermion, measured by t$_3$, and the rotation charge, $Q_r$, of the trefoil with the hypercharge of the standard fermion, measured by $t_0$.  We shall then assume that the elementary fermions are quantum trefoils and that their total electric charge may be written as either
\newpage
\begin{equation}
Q_e=Q_w+Q_r
\end{equation}
\begin{center}
or
\end{center}
\begin{equation}
Q_e=e(t_3+t_0)
\end{equation}

\noindent similar to the way that their total angular momentum and magnetic moment may be written as the sum of the spin and orbital contributions. (The analogy may be carried further since the spin and writhe are both localized: the spin is localized on the particles, and the writhe is localized at the crossings, while the orbital angular momentum describes the entire orbital motion, and the knot rotation is computed for the entire knot.)

From Table 2.1 one may also read the following relation between the quantum trefoils of the knot model, as measured by $(N, w, r)$, and the fermions of the standard model, as described by the isotopic charge and hypercharge:$^{(3)(4)}$
\begin{equation}
(N, w,r+1) = 6(t, -t_3, -t_0)
\end{equation}
\noindent or by (1.7)
\begin{equation}
(j, m, m') = 3(t, -t_3,-t_0)
\end{equation}

The empirical correspondence between the topological description of the four quantum trefoils and the quantum numbers of the four families of fermions is encapsulated in (2.14). Otherwise stated, there is a unique way of satisfying (2.14) with the four quantum trefoils and the four classes of fermions.

\section{The Fermion-Boson Interaction in the Knot Model}

According to the rule (2.1) the Fermion-Boson interaction terms of the standard action are multiplied by the form factors
\begin{equation}
\overline{\mbox{D}}^{j''}_{m''p''} \mbox{D}^j_{mp} \mbox{D}^{j'}_{m'p'}
\end{equation}

\noindent in passing from the standard model to the knot model.  Here D$^{j'}_{m'p'}$ and $\overline{\mbox{D}}^{j''}_{m''p''}$ multiply the initial and final Fermi operators, resp., while D$^j_{mp}$ multiplies the mediating boson operator of the standard model.

By U$_m$(1) $\times$ U$_{p}$(1) invariance of (3.1) we have
\begin{equation}
(m,p)=(m'',p'')-(m',p')
\end{equation}

\noindent By (2.15) the empirical relations
\begin{eqnarray}
(m',p') &=& -3(t_3,t_0)'  \nonumber \\
(m'',p'') &=& -3(t_3,t_0)''
\end{eqnarray}
\noindent hold for the fermion operators.

Then by (3.2) and (3.3)
\begin{equation}
(m,p)=-3[(t_3,t_0)''-(t_3,t_0)']
\end{equation}

\noindent In passing from the standard model to the knotted model we retain SU(2)$\times$U(1) invariance and therefore the conservation of $t_3$ and $t_0$ separately: $t_3''=t_3'+t_3$ and $t_0''=t_0'+t_0$. The conservation of $t_3$ and $t_0$ is also a consequence of the required U$_a$(1)$\times$U$_b$(1) invariance of the action, and is expressed by the separate conservation of the writhe and rotation charges.

Then by (3.4)
\begin{equation}
(m,p)=-3(t_3,t_0)
\end{equation}

\noindent for the intermediate boson as well as for the initial and final fermions.  Also since
\begin{equation}
j'+j'' \ge j \ge |j'-j''| \hskip0.4cm and \hskip0.4cm j'=j'' = \frac{3}{2}
\end{equation}
\noindent $j$ is fixed by
\begin{equation}
3 \ge j \ge |m|
\end{equation}
\noindent Then one has by $\left( 3.5 \right)$
\begin{equation}
\mbox{D}^j_{mp} = \mbox{D}^3_{\pm 3 0}
\end{equation}
\noindent for the charged vector bosons, where $\left( t, t_3, t_0 \right) = \left(1, \pm 1, 0 \right)$ and we set $j=3$. We assume a similar relation for the neutral vector boson where $\left( t, t_3, t_0 \right) = \left( 1, 0, 0 \right)$.

Hence there is an empirical basis, dependent also on the postulated symmetries, for
\begin{equation}
(j,m,m')=3(t,-t_3,-t_0)
\end{equation}
\noindent for both the fermions and vector bosons of the knotted model.

In both cases one may write for the field operator of the knot model
\begin{eqnarray}
\hat{\Psi}(t,t_3,t_0) &=& \Psi(t,t_3,t_0)\mbox{D}^j_{mm'} \nonumber \\
&=& \Psi \left( t, t_3, t_0 \right) \mbox{D}^{3t}_{-3t_3 - 3t_0}
\end{eqnarray}

\noindent where $\Psi(t,t_3,t_0)$ is the field operator of the standard model and in both cases we have (3.9).

Here $\Psi$ means left chiral when it refers to the elementary fermion. As in the standard model, we assume that the right chiral field is an isotopic singlet, but in the knot extension we assume it has the same knot factor as its left chiral partner. The right chiral state does not satisfy (3.10).

\section{The Preon Representations$^{(3)(4)}$}
In the model that we are describing, the elementary fermions, with $t=\frac{1}{2}$, are the simplest quantum knots, the trefoils, with $N=3$ and by (1.7)
\begin{equation}
j=\frac{1}{2}N=\frac{3}{2}(=3t)
\end{equation}

\noindent In the same model the electroweak bosons, with $t=1$, are quantum ditrefoils, with $N=6$ and
\begin{equation}
j=\frac{1}{2}N=\frac{6}{2}(=3t)
\end{equation}

\noindent Then the elementary fermions lie in the $j=\frac{3}{2}$ representation while the electroweak bosons lie in the $j=3$ representation of SLq(2).

We now consider the adjoint ($j=1$) representation and the fundamental ($j=\frac{1}{2}$) representation of SLq(2) as defined by (1.2).  After dropping the $\mbox{A}^j_{mm'}$ these are shown below in tables 4.1 and 4.2.

\begin{center}
$\underline{\bf{Table \hspace{0.1cm}4.1}}$
\end{center}
 \begin{center}
 D$^{\frac{1}{2}}_{mm'}$: \hspace{0.7cm}
 \begin{tabular}{c | c c}
 \backslashbox[-20pt][l]{$m$}{$m'$} & $\frac{1}{2}$ & -$\frac{1}{2}$\\
 \hline
 $\frac{1}{2}$ & a & b\\
 -$\frac{1}{2}$ & c & d
 \end{tabular}
 \end{center}
 
 \newpage
 
 \begin{center}
 $\underline{\bf{Table \hspace{0.1cm} 4.2}}$
 \end{center}
\begin{center}
 D$^1_{mm'}$: \hspace{0.7cm}
 \begin{tabular}{c | c c c}
 \backslashbox[-10pt][l]{$m$}{$m'$} & 1 & 0 & -1\\
 \hline
 1 & $a^2$ & $ab$ & $b^2$\\
 0 & $ac$ & $ad+bc$ & $bd$\\
 -1 & $c^2$ & $cd$ & $d^2$
 \end{tabular}
 \end{center}
 
 \noindent We shall refer to the members of the D$^{\frac{1}{2}}$ and D$^1$ representations as fermionic and bosonic preons respectively.
 
 To determine ($t_3,t_0,Q$) for the fermionic and bosonic preons we shall extend the relations empirically established for the elementary fermions, then extended to the electroweak bosons, and generally expressed in D$^{3t}_{-3t_3-3t_0}$.
 \noindent The results for preons are shown in tables (4.3) and (4.4) where eqns. (2.15) and (2.11), namely:
 \begin{displaymath}
 (t,t_3,t_0) = \frac{1}{3} (j,-m,-m') \tag{2.15} 
 \end{displaymath}
 \begin{displaymath}
 Q = \nonumber e (t_3 + t_0) = -\frac{e}{3}(m+m') \tag{2.11}
 \end{displaymath}
 
 \noindent are now read from right to left.

 \begin{center}
 \underline{\bf{Table 4.3}}
 \end{center}
 \begin{center}
 \bf{Fermionic Preons $t=\frac{1}{6}$}
 \end{center}
\begin{center}
\begin{tabular}{c | c c c c}
 & $t$ & $t_3$ & $t_0$ & $Q$ \\
 \hline
 $a$ & $\frac{1}{6}$ & -$\frac{1}{6}$ & -$\frac{1}{6}$ & -$\frac{e}{3}$\\
 $b$ & $\frac{1}{6}$ & -$\frac{1}{6}$ & $\frac{1}{6}$ & 0\\
 $d$ & $\frac{1}{6}$ & $\frac{1}{6}$ & $\frac{1}{6}$ & $\frac{e}{3}$\\
 $c$ & $\frac{1}{6}$ & $\frac{1}{6}$ & -$\frac{1}{6}$ & 0\\
 \end{tabular}
 \end{center}
 \newpage
 \begin{center}
 \underline{\bf{Table 4.4}}
 \end{center}
 \begin{center}
 \bf{Bosonic Preons $t=\frac{1}{3}$}
 \end{center}
\begin{center}
\begin{tabular}{c | c c c c || c | c c c c || c | c c c c}
& $t_3$ & $t_0$ & $\frac{Q}{e}$ & $\mbox{D}^1_{mm'}$ & & $t_3$ & $t_0$ & $\frac{Q}{e}$ & $\mbox{D}^1_{mm'}$ & & $t_3$ & $t_0$ & $\frac{Q}{e}$ & $\mbox{D}^1_{mm'}$ \\
\hline
$\mbox{D}^1_{11}$ & -$\frac{1}{3}$ & -$\frac{1}{3}$ &-$\frac{2}{3}$ & $a^2$ & $\mbox{D}^1_{01}$ & 0 & -$\frac{1}{3}$ & -$\frac{1}{3}$ & $ac$ & $\mbox{D}^1_{-11}$ & $\frac{1}{3}$ & -$\frac{1}{3}$ & 0 & $c^2$\\
$\mbox{D}^1_{10}$ & -$\frac{1}{3}$ & 0 & -$\frac{1}{3}$ & $ab$ & $\mbox{D}^1_{00}$ & 0 & 0 & 0 & $ad + bc$ & $\mbox{D}^1_{-10}$ & $\frac{1}{3}$ & 0 & $\frac{1}{3}$ & $cd$\\
$\mbox{D}^1_{1-1}$ & -$\frac{1}{3}$ & $\frac{1}{3}$ & 0 & $b^2$ & $\mbox{D}^1_{0-1}$ & 0 & $\frac{1}{3}$ & $\frac{1}{3}$ & $bd$ & $\mbox{D}^1_{-1-1}$ & $\frac{1}{3}$ & $\frac{1}{3}$ & $\frac{2}{3}$ & $d^2$\\
\end{tabular}
\end{center}

\noindent The values of $(t, t_3, t_0)$ in these tables have meaning for the knot model but not for isotopic spin.  In this respect the knot model provides an extension of the isotopic spin. The fractional values of $t_3$ and $t_0$ follow from (2.15) and measure the writhe and rotation charges, respectively.

According to Table (4.3) there are two preons, a and b, charged and neutral respectively and their respective anti-particles, d and c, with opposite charge and hypercharge. These preons agree with the preons proposed by Harari and Shupe.$^{(13)}$ We may also regard $a$ and $c$ and $d$ and $b$ as belonging to $t_3$-doublets.

We now show that all particles belonging to higher representations may be regarded as built up out of preons (a, b, c, d) insofar as the values of $(t_3, t_0, Q)$ for all the composite particles may be obtained by adding the $(t_3, t_0, Q)$ of each of the constituent preons.

We have in general by (1.2)
\begin{equation}
\mbox{D}^j_{mm'}(a,b,c,d)=\sum_{\begin{array}{rllll} 0 & \le & s & \le &  n_+ \\ 0 & \le & t & \le & n_- \end{array}} \mbox{A}^j_{mm'}(q,s,t) \delta (s+t,n'_+) a^s b^{n_+-s} c^t d^{n_--t}
\end{equation}

Denote the exponents of $(a, b, c, d)$ by $(n_a, n_b, n_c, n_d)$.  These will vary from term to term but there are the following structural constraints on the sum (4.3)
\begin{eqnarray}
n_a + n_b + n_c + n_d &=& 2j \\
n_a + n_b - n_c -n_d &=& 2m \\
n_a - n_b + n_c - n_d &=& 2m'
\end{eqnarray}

\noindent But by (2.14) and (2.15)
\begin{equation}
(j, m, m') = 3(t, -t_3, -t_0)
\end{equation}

\noindent and
\begin{equation}
(j, m, m') = \frac{1}{2}(N, w,r+o)
\end{equation}

\noindent Eqns (4.7) and (4.8) are the basic empirical and topological constraints defining the knot model. We have shown how they hold for the $j=3/2$ and $j=3$ representations. We now assume that they hold for all representations that we consider.

We may now rewrite the structural equations (4.4)-(4.6) in terms of $(t, t_3, t_0)$ or alternatively in terms of $(N, w, r)$.

We shall also retain
\begin{eqnarray}
Q &=& e(t_3 + t_0) \nonumber\\
&=& -\frac{e}{3}(m+m') \nonumber
\end{eqnarray}
\noindent for all representations.

In terms of $(t, t_3, t_0, Q)$ Equations (4.4), (4.5), and (4.6) become by (4.7)
\begin{eqnarray}
t &=& \frac{1}{6} (n_a + n_b + n_c + n_d) \\
t_3 &=& -\frac{1}{6} (n_a + n_b - n_c - n_d) \\
t_0 &=& -\frac{1}{6}(n_a - n_b + n_c - n_d)
\end{eqnarray}

\noindent Then
\begin{equation}
Q = e(t_3 + t_0) = -\frac{e}{3}(n_a - n_d)
\end{equation}
\noindent By Table (4.3) the equations (4.9), (4.10), (4.11), and (4.12) may be written as follows
\begin{eqnarray}
t &=& n_at_a + n_bt_b + n_ct_c +n_dt_d \\ 
t_3 &=& n_a(t_3)_a + n_b(t_3)_b + n_c(t_3)_c + n_d(t_3)_d \\
t_0 &=& n_a(t_0)_a + n_b(t_0)_b + n_c(t_0)_c + n_d(t_0)_d \\
Q &=& n_aQ_a + n_bQ_b + n_cQ_c + n_dQ_d
\end{eqnarray}

\noindent or 
\begin{equation}
(t, t_3, t_0, Q) = \sum_{p=(a, b, c, d)}n_p(t_p, t_{3p}, t_{0p}, Q_p)
\end{equation}

If we now interpret $(a, b, c, d)$ as the creation operators for the $(a, b, c, d)$ preons, then the $(n_a, n_b, n_c, n_d)$ represent the number of $(a, b, c, d)$ preons respectively in each term.  Then (4.17) states that the composite particle on the left with quantum numbers $(t, t_3, t_0, Q)$ may be regarded as a superposition of separate states, all of which have the same $(t, t_3, t_0, Q)$ but contain different numbers of preons $(n_a, n_b, n_c, n_d)$ with quantum numbers $(t_p, t_{3p}, t_{0p}, Q_p)$ where $p = (a, b, c, d)$.

We illustrate (4.13), (4.14), (4.15), and (4.16) in the following Tables 4.5, 4.6, and 4.7.

\noindent These tables may be read in two ways:
\begin{quote}
(a) as describing creation operators representing the internal state of a composite particle, or
\end{quote}
\begin{quote}
(b) as describing a product of creation operators for the component preons.
\end{quote}

\begin{center}
\bf{Table 4.5}
\end{center}
\begin{center}
\bf{\underline{Preons ($j$ = $\frac{1}{2}$)}}
\end{center}
\begin{center}
\begin{tabular}{c | c c c c}
 & Q & $t_3$ &  $t_0$ & $\mbox{D}^{3t}_{-3t_3-3t_0}$ \\
 \hline
 a & -$\frac{e}{3}$ & -$\frac{1}{6}$ & -$\frac{1}{6}$ & $\mbox{D}^{\frac{1}{2}}_{\frac{1}{2} \frac{1}{2}} \sim a$ \\
 b & 0 & -$\frac{1}{6}$ & $\frac{1}{6} $ & $\mbox{D}^{\frac{1}{2}}_{\frac{1}{2} -\frac{1}{2}} \sim b$\\
 c & 0 & $\frac{1}{6}$ & -$\frac{1}{6}$ & $\mbox{D}^{\frac{1}{2}}_{-\frac{1}{2} \frac{1}{2}} \sim c$\\
 d & $\frac{e}{3}$ & $\frac{1}{6}$ & $\frac{1}{6}$ & $\mbox{D}^{\frac{1}{2}}_{-\frac{1}{2} -\frac{1}{2}} \sim d$\\
\end{tabular}
\end{center}

\vspace{1.0 cm}
\begin{center}
\bf{Table 4.6}
\end{center}
\begin{center}
\bf{\underline{Fermions ($j$= $\frac{3}{2}$)}}
\end{center}
\begin{center}
\begin{tabular}{c | c c c c}
 & Q & $t_3$ & $t_0$ & $\mbox{D}^{3t}_{-3t_3-3t_0}$\\
 \hline
 $l$ & -$e$ & -$\frac{1}{2}$ & -$\frac{1}{2}$ & $\mbox{D}^{\frac{3}{2}}_{\frac{3}{2} \frac{3}{2}} \sim a^3$\\
 $\nu$ & 0 & $\frac{1}{2}$ & -$\frac{1}{2}$ & $\mbox{D}^{\frac{3}{2}}_{-\frac{3}{2} \frac{3}{2}} \sim c^3$\\
 d & -$\frac{1}{3}e$ & -$\frac{1}{2}$ & $\frac{1}{6}$ & $\mbox{D}^{\frac{3}{2}}_{\frac{3}{2} -\frac{1}{2}} \sim ab^2$\\
 $u$ & $\frac{2}{3}e$ & $\frac{1}{2}$ & $\frac{1}{6}$ & $\mbox{D}^{\frac{3}{2}}_{-\frac{3}{2}  -\frac{1}{2}} \sim cd^2$\\
 \end{tabular}
\end{center}

\vspace{1.0cm}
\begin{center}
\bf{Table 4.7}
\end{center}
\begin{center}
\bf{\underline{Electroweak Vectors ($j$ = 3)}}
\end{center}
\begin{center}
\begin{tabular}{c | c c c c c}
 & Q & $t$ & $t_3$ & $t_0$ & $\mbox{D}^{3t}_{-3t_3-3t_0}$ \\
 \hline
 W$^+$ & $e$ & 1 & 1 & 0 & $\mbox{D}^{3}_{-30} \sim c^3d^3$\\
 W$^-$ & -$e$ & 1 & -1 & 0 & $\mbox{D}^{3}_{30} \sim a^3b^3$\\
 W$^3$ & 0 & 1 & 0 & 0 & $\mbox{D}^{3}_{00} \sim f_3(b c)$ \\
 W$^0$ & 0 & 0 & 0 & 0 & $\mbox{D}^{0}_{00} \sim f_0(b c)$ \\
\end{tabular}
\end{center}

\noindent The tables (4.5), (4.6), and (4.7) are computed by $\left(4.3 \right)$ after dropping the A$^j_{mm'}$.

\noindent The field operators are now expanded in the complete polynomials $\mbox{D}^j_{mm'}$ expressed in terms of the preon operators $(a,b,c,d)$. All terms in these polynomials have the same charge and hypercharge as the composite particle on the left side of $(1.2)$. If $\mbox{D}^j_{mm'}$ is a monomial (like the elementary fermions) the field operator creates a single state, but otherwise it creates a superposition of several states.

\noindent According to Tables 4.5 and 4.6 the leptons are composed of three $a$-preons while the neutrinos are composed of three $c$-preons. The down quarks contain one $a$ and two $b$-preons while the up quarks contain one $c$ and two $d$-preons.

\noindent These descriptions of the elementary fermions as three preon structures are in agreement with the Harari-Shupe model.$^{(13)}$ In Table 4.7 the charged vectors are also in agreement with the same model, but the neutral vector $\mbox{W}_{\mu}^3$ is the superposition of four states of six preons given by
\begin{displaymath}
\mbox{D}_{00}^3=\mbox{A}(0,3)b^3c^3 + \mbox{A}(1,2)ab^2c^2d + \mbox{A}(2,1)a^2bcd^2 + \mbox{A}(3,0) a^3d^3
\end{displaymath}
according to (1.2) and expressible as a function of the neutral operator $bc$.

\newpage

\section{The Complementary Models$^{(4)}$}
\noindent Since ($N, w, r+o$) = 2($j, m, m'$), equations (4.4)-(4.6) giving ($j, m, m'$) may also be read as knot relations as follows:
\begin{eqnarray}
N &=& n_a + n_b + n_c + n_d \\
w &=& n_a + n_b - n_c - n_d \\
r + o &=& n_a + n_c - n_b - n_d
\end{eqnarray}

\noindent There are only three equations to determine the four $(n_a, n_b, n_c, n_d)$.  Therefore the composite particle, either $(t, t_3, t_0)$ or $(N, w, r)$, is in general a superposition of several components with different sets of $(n_a, n_b, n_c, n_d)$.

\noindent Equation (5.1) states that the total number of preons equals the number of crossings ($N$).

\noindent Since we assume that the preons are fermions, the knot describes a fermion or a boson depending on whether the number of crossings is odd or even.

\noindent The meaning of equations (5.2) and (5.3) becomes clear if we note that $a$ and $d$ are antiparticles with opposite charge and hypercharge, while $b$ and $c$ are neutral antiparticles with opposite values of the hypercharge.

\noindent We may therefore introduce the preon numbers
\begin{eqnarray}
\nu_a &=& n_a - n_d \\
\nu_b &=& n_b - n_c
\end{eqnarray}

\noindent Then (5.2) and (5.3) may be rewritten as
\begin{eqnarray}
\nu_a + \nu_b &=& w = -6t_3 \\
\nu_a - \nu_b &=& r + o = -6t_0
\end{eqnarray}

\noindent By (5.6) and (5.7) the conservation of the preon numbers and of charge and hypercharge is equivalent to the conservation of the writhe and rotation which are topologically conserved at the classical level.  In this respect, these conservation laws may be regarded as topological.

The SLq(2) equations (5.1), (5.2), (5.3) hold for all representations and therefore for preons as well as for knots, although the preons are twisted loops rather than knots.  If the indices $(N, w, r)$ for the fermionic preons are determined in the same way as for knots, one finds $N=1$, $w=\pm 1$, and $r=0$.  Then by (5.3) the odd integer, $o$, is for preons
\begin{equation}
o=n_a + n_b - n_c - n_d
\end{equation}
It follows that $o=1$ for $a$ and $b$, and that $o=-1$ for the antiparticles, $d$ and $c$.

Viewed as a knot, a fermion becomes a boson when the number of crossings is changed by attaching or removing a curl.  This picture is consistent with the view of a curl as an opened preon loop.

Corresponding to the representations of the four elementary fermions as three preon states, there is the complementary representation of the four trefoils as composed of three overlapping preon loops as shown in Fig. 5.1.
\newpage

\begin{center}
\underline{\textbf{Fig. 5.1}}
\end{center}
\begin{center}
\begin{tabular}{cc|cc}
 & \underline{$\left( w, r, o \right)$} & & \underline{$\left( w, r, o \right)$} \\
 Leptons, $\mbox{D}^{\frac{3}{2}}_{\frac{3}{2} \frac{3}{2}} \sim a^3$ & & $a$-preons, $\mbox{D}^{\frac{1}{2}}_{\frac{1}{2} \frac{1}{2}}$ \\
 
 \includegraphics[scale=0.45]{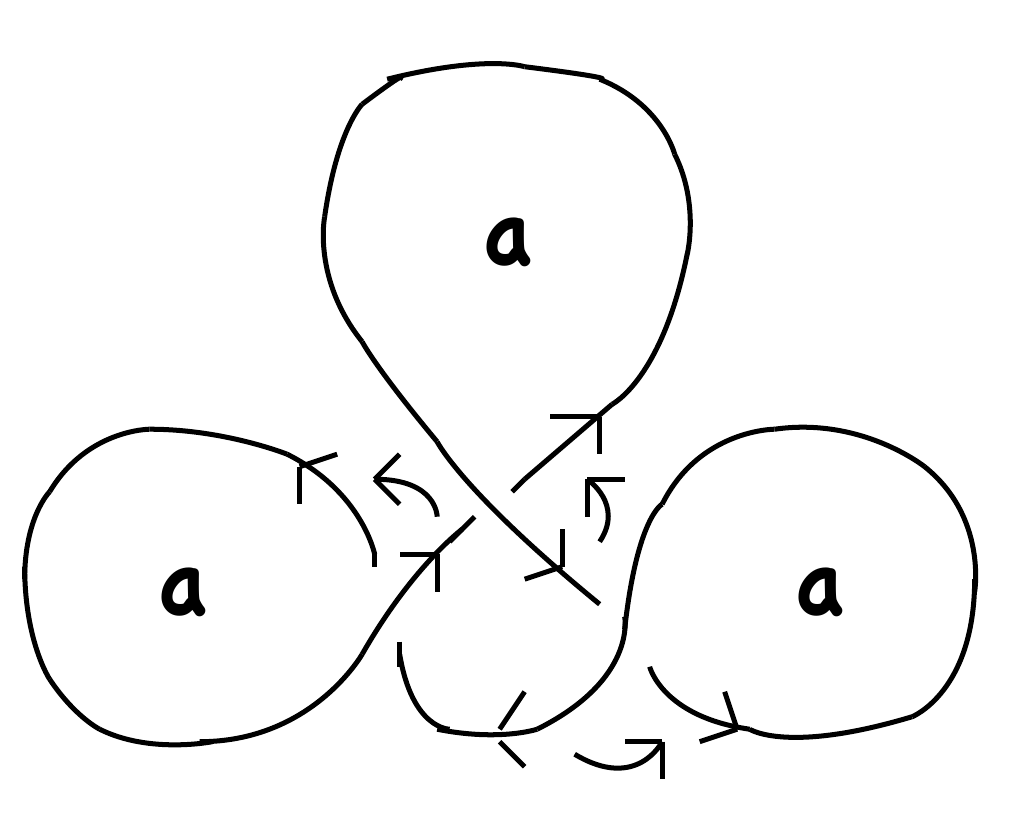} & $\left(3,2,1\right)$ & \includegraphics[scale=0.45]{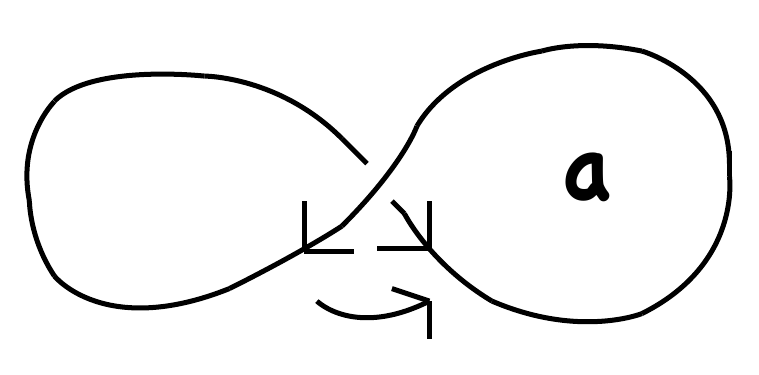} & $\left(1,0,1\right)$ \\
 \hline
 Neutrinos, $\mbox{D}^{\frac{3}{2}}_{-\frac{3}{2} \frac{3}{2}} \sim c^3$ & & $c$-preons, $\mbox{D}^{\frac{1}{2}}_{-\frac{1}{2} \frac{1}{2}}$ & \\
 
 \includegraphics[scale=0.45]{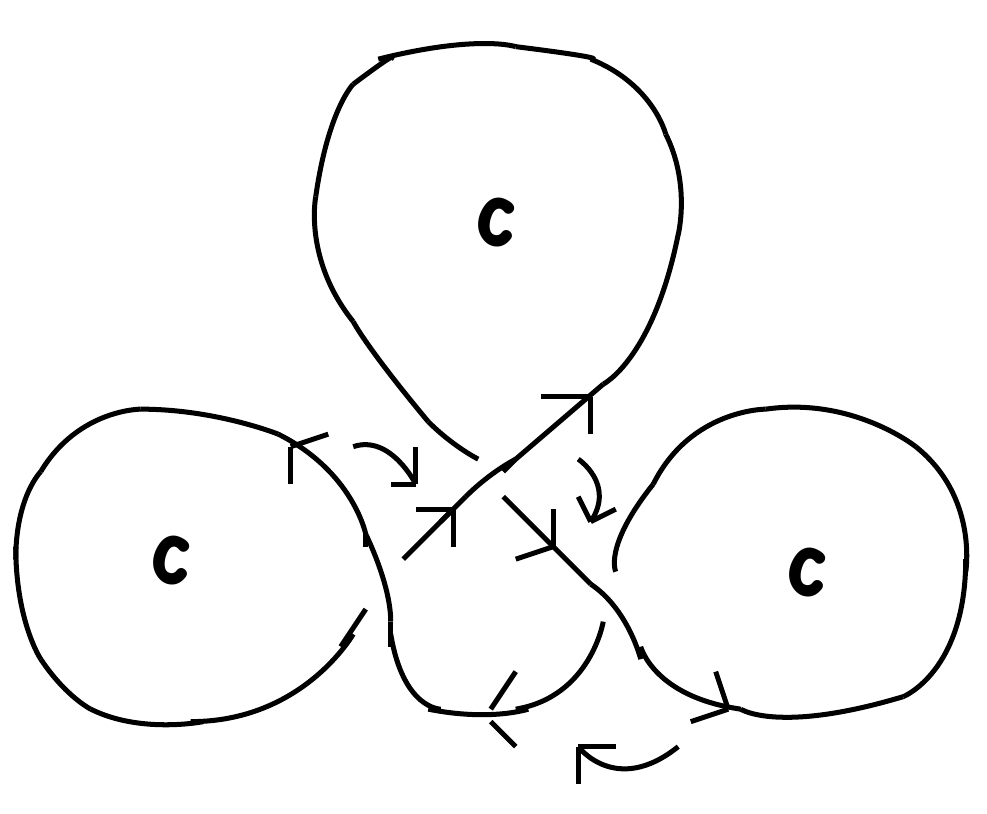} & $\left(-3,2,1 \right)$ & \includegraphics[scale=0.45]{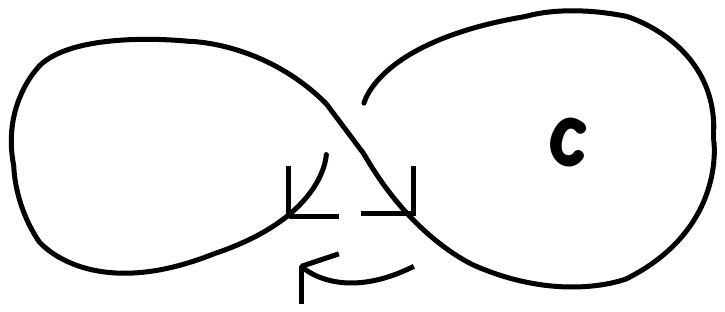} & $\left(-1,0,1 \right)$\\
 \hline
 $d$-quarks, $\mbox{D}^{\frac{3}{2}}_{\frac{3}{2} -\frac{1}{2}} \sim ab^2$ & & $b$-preons, $\mbox{D}^{\frac{1}{2}}_{\frac{1}{2}-\frac{1}{2}}$ & \\
 
\includegraphics[scale=0.45]{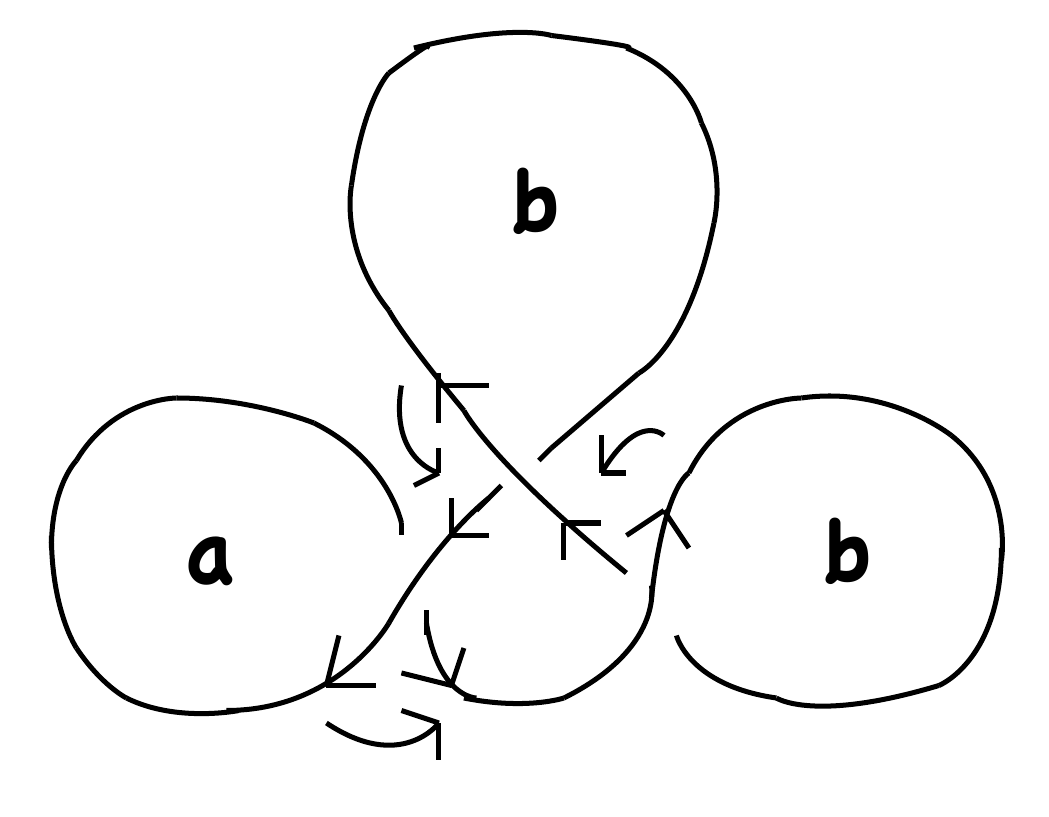} & $\left( 3,-2,1\right)$ & \includegraphics[scale=0.45]{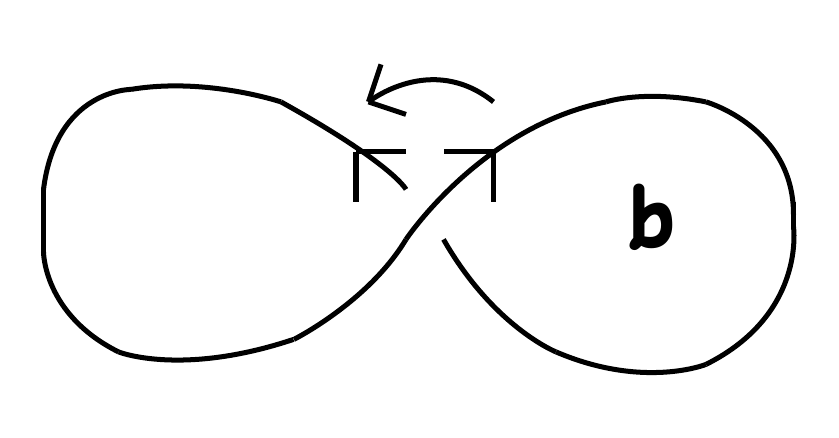} & $\left(1,0,-1\right)$ \\
\hline
$u$-quarks, $\mbox{D}^{\frac{3}{2}}_{-\frac{3}{2} -\frac{1}{2}} \sim cd^2$ & & $d$-preons, $\mbox{D}^{\frac{1}{2}}_{-\frac{1}{2} -\frac{1}{2}}$ \\

\includegraphics[scale=0.45]{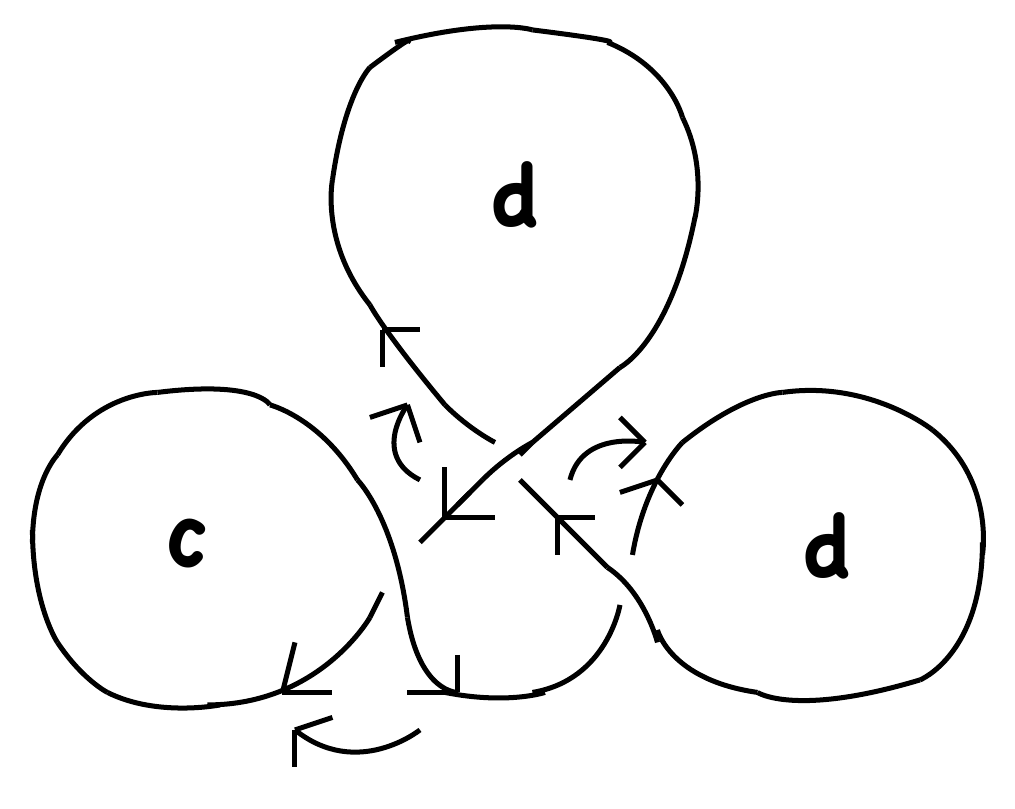} & $\left(-3,-2,1 \right)$ & \includegraphics[scale=0.45]{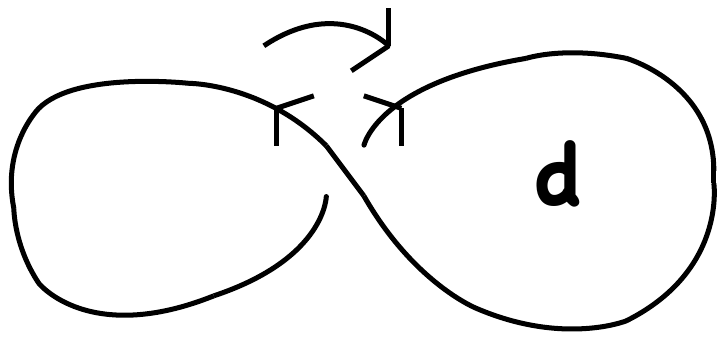} & $\left(-1,0,-1 \right)$ \\
\end{tabular}
\end{center}
\begin{center}
$Q = -\frac{e}{6} \left( w+r+o \right)$ \\
$\left(j,m,m' \right) = \frac{1}{2} \left(N, w, r+o \right)$
\end{center}
\newpage


In interpreting Fig. 5.1 note that the two lobes of all the preons make opposite contributions to the rotation, $r$, so that the total rotation of each preon vanishes.  When the three $a$-preons and $c$-preons are combined to form leptons and neutrinos, respectively, each of the three labelled circuits is counterclockwise and contributes $+1$ to the rotation while the single unlabelled shared circuit is clockwise and contributes $-1$ to the rotation so that the total $r$ for both leptons and neutrinos is $+2$.  For the quarks the three labelled loops contribute $-1$ and the shared loop $+1$ so that $r=-2$.

\noindent Written in terms of $(N, w, r)$ and $(N, w, r)_p$ the equations describing the composite particles are
\begin{equation}
\tag{$5.9N$}
N = \sum_p n_p N_p 
\end{equation}
\begin{equation}
\tag{$5.9w$}
w = \sum_p n_p w_p
\end{equation}
\begin{equation}
\tag{$5.9\tilde{r}$}
\tilde{r} = \sum_p n_p \tilde{r}_p
\end{equation}

\noindent where $p = (a,b,c,d)$ and 
\begin{equation}
\setcounter{equation}{10}
\tilde{r} = r + o
\end{equation}
For preons
\begin{equation}
\tilde{r}_p = o_p
\end{equation}
For the elementary fermions of the standard model
\begin{equation}
\tilde{r} = r +1
\end{equation}

\noindent These considerations lead one to view the symmetry of an elementary particle, defined by representations of the SLq(2) algebra, in any of the following ways:
\begin{equation}
\mbox{D}^j_{mm'} = \mbox{D}^{3t}_{-3t_3-3t_0} = \mbox{D}^{\frac{N}{2}}_{\frac{w}{2} \frac{\tilde{r}}{2}} = \tilde{\mbox{D}}^{N'}_{\nu_a \nu_b}
\end{equation}

\noindent where $N'$ is the total number of preons.  The quantum knot-preon complementary representations are related by
\begin{equation}
\tilde{\mbox{D}}^{N'}_{\nu_a \nu_b} = \sum_{Nwr} \delta (N', N) \delta(\nu_a + \nu_b, w) \delta(\nu_a - \nu_b, \tilde{r}) \mbox{D}^{\frac{N}{2}}_{\frac{w}{2} \frac{\tilde{r}}{2}}
\end{equation}

\section{Gluon Charge$^{(4)}$}

The previous considerations are based on electroweak physics. To describe the strong interactions it is necessary according to the standard model to introduce SU(3). We may assume that only the $a$ and $c$ preons carry gluon charge and that the $b$ and $d$ preons are color singlets.$^{(12)}$ The $a$ and $c$ preon operators then appear in triplicate as $(a_i, c_i)$ where $i=(R,Y,G)$ without changing the $(a, b, c, d)$ algebra. These colored preon operators provide a basis for the fundamental representation of SU(3) just as the colored quark operators do in standard theory.  To adapt the electroweak operators to the requirements of gluon fields we make the following replacements:


\begin{center}
leptons: $a^3$ $\rightarrow$ $\varepsilon^{ijk}a_ia_ja_k$
\end{center}
\begin{center}
neutrinos: $c^3$ $\rightarrow$ $\varepsilon^{ijk}c_ic_jc_k$
\end{center}
\begin{center}
down quarks: $ab^2$ $\rightarrow$ $a_ib^2$
\end{center}
\begin{center}
up quarks: $cd^2$ $\rightarrow$ $c_i d^2$
\end{center}

\noindent Here $(a_i, c_i)$ are creation operators for colored preons.  Then the leptons and neutrinos are color singlets while the quark states correspond to the fundamental representations of SU(3), as required by the standard model.


\vskip1.4cm
\centerline{\bf{\Large{II. Dynamics}}}

\section{The Knot Lagrangian}

The Lagrangian of the standard model at the electroweak level may be written as follows
\begin{equation}
\mathcal{L}_{st} = -\frac{1}{4}\mbox{Tr}\mbox{W}^{\mu \lambda}\mbox{W}_{\mu \lambda} - \frac{1}{4} \mbox{H}^{\mu \lambda}\mbox{H}_{\mu \lambda} + i[\bar{L}\nabla L + \bar{R} \nabla R] + \frac{1}{2} \overline{\nabla\varphi} \nabla\varphi - V(\bar{\varphi}\varphi) - \frac{m}{\rho}[\bar{L}\varphi R + \bar{R} \bar{\varphi} L]
\end{equation}

\noindent To obtain the knot Lagrangian we attempt to replace every field operator of the standard model by
\begin{equation}
\Psi \rightarrow \Psi \mbox{D}^{j}_{mm'}(\Psi)
\end{equation}

\noindent where the $\mbox{D}^{j}_{mm'}(\Psi)$ are determined empirically, as discussed in part I, subject to the requirement that every term of the modified Lagrangian be SU(2)$\times$U(1) and U$_a$(1)$\times$U$_b$(1) invariant.

To implement (7.2) we begin with
\begin{equation}
\Psi(t, t_3, t_0) \rightarrow \Psi(t, t_3, t_0)\mbox{D}^{3t}_{-3t_3-3t_0}
\end{equation}
\noindent and
\begin{equation}
Q = e(t_3 + t_0)
\end{equation}
\noindent
or
\begin{equation}
Q = -\frac{e}{3}(m + m') = -\frac{e}{6}(w + r+1)
\end{equation}

\noindent for the left chiral field, L, and for the vector bosons as discussed in part I.

We assume that every right chiral field, R, has the same knot factor, $\mbox{D}^{j}_{mm'}$, as the corresponding L-field.  We shall also assume that R is an isotopic singlet, with $t=0$, here as in the standard model.  Then R does not and is not required to satisfy (7.3).  Since we shall, however, assume that
\begin{equation}
Q=-\frac{e}{3}\left(m+m'\right)
\end{equation}
holds for both L and R, it follows that L and R carry the same electric charge.


If the modification of the standard model is made according to the preceding substitutions, it will be shown that the new Lagrangian will be U$_a \left(1 \right) \times$ U$_b \left(1 \right)$ invariant as required, and all new factors appearing in the new Lagrangian will be functions of $bc$.

To obtain the knot Lagrangian the standard Lagrangian will be replaced term by term beginning with the mass terms.

\section{The Mass Terms}

In the standard model L and $\varphi$ are isotopic doublets while ($\bar{L} \varphi$) and R are isotopic singlets.  We retain this isotopic structure and continue to follow the standard model by going to the unitary gauge where $\varphi$ has a single component which is neutral.  In passing to the SLq(2) algebra we assume that $\varphi$ is a SLq(2) singlet.

To obtain the mass term for the leptons, we write for the left chiral fields

\begin{equation}
L(l) =  \begin{pmatrix}
\nu_L & \mbox{D}^{\frac{3}{2}}_{-\frac{3}{2} \frac{3}{2}}\\
l_L & \mbox{D}^{\frac{3}{2}}_{\frac{3}{2} \frac{3}{2}}
\end{pmatrix} \to
\begin{pmatrix}
\nu_Lc^3\\
l_La^3 \end{pmatrix}
\end{equation}

\noindent
where $ \begin{pmatrix} \nu_L\\ l_L \end{pmatrix}$ is the corresponding doublet of the standard model.  The numerical coefficients in $\mbox{D}^{j}_{mm'}$ have been temporarily dropped and the monomials $\mbox{D}^{\frac{3}{2}}_{-\frac{3}{2}\frac{3}{2}}$ and $\mbox{D}^{\frac{3}{2}}_{\frac{3}{2} \frac{3}{2}}$ have been replaced by $c^3$ and $a^3$.  Now, having assumed that $\varphi$ is a SLq(2) singlet and that the knot factors for R and L are the same, one has
\begin{equation}
\bar{L}(l)\varphi_l R(l) = \left[ \begin{pmatrix}
\bar{\nu}_L\bar{c}^3 & \bar{l}_L \bar{a}^3
 \end{pmatrix} \cdot \begin{pmatrix}
 0\\
 \rho_l
 \end{pmatrix}
 \right] (l_Ra^3)
\end{equation}

\noindent Here the Higgs doublet is in the unitary gauge and $\rho_l$ is its neutral component.

\noindent Then
\begin{equation}
\bar{L}(l)\varphi_lR(l) = \rho_l (\bar{a}^3 a^3) \cdot \bar{l}_L l_R
\end{equation}

\noindent Similarly
\begin{eqnarray}
\bar{R}(l)\bar{\varphi}_lL(l) = (\bar{a}^3\bar{l}_R) \cdot \left[  \begin{pmatrix} 0 & \rho_l \end{pmatrix} \cdot  \begin{pmatrix} \nu_Lc^3\\ l_La^3 \end{pmatrix} \right]\\
= \rho_l (\bar{a}^3 a^3) \bar{l}_Rl_L
\end{eqnarray}

\noindent Hence
\begin{equation}
\bar{L}(l)\varphi_lR(l) + \bar{R}(l)\bar{\varphi}_lL(l) = \rho_l(\bar{a}^3 a^3)(\bar{l}_Ll_R + \bar{l}_Rl_L)
\end{equation}
\noindent and
\begin{eqnarray}
\bra{n}\bar{L}(l)\varphi_lR(l) + \bar{R}(l)\bar{\varphi}_lL(l)\ket{n} = \rho_l \bra{n}\bar{a}^3a^3\ket{n}(\bar{l}l)\\
= m_l \bar{l}l \nonumber
\end{eqnarray}

\noindent where
\begin{equation}
m_l = \rho_l \bra{n} \bar{a}^3 a^3 \ket{n}
\end{equation}

\noindent and $\rho_l$ is the vacuum expectation value of the Higgs that fixes the lepton masses (multiplied by numerical factors dropped in (8.1)).

In (8.8) $\bar{a}^3a^3$ is an operator holding for any member of the lepton family and by the algebra $\left( 1.1 \right)$ is expressible as a simple polynomial in $bc$.  We shall distinguish the three mass states by $\ket{n}$, $n=0, 1, 2,$ where $\ket{n}$ is an eigenstate of $b$ and $c$ and of the mass operator, expressed as a function of $b$ and $c$.  We therefore replace the lepton contribution to the mass term of the standard model by
\begin{equation}
\sum_n \bra{n}\bar{L}_l \varphi_lR_l + \bar{R}_l \bar{\varphi}_lL_l \ket{n}
\end{equation}

\noindent where $n$ is summed over the three generations of leptons.  Since $\bra{n} \bar{a}^3 a^3 \ket{n}$ depends on $n$ while $\rho_l$ does not depend on $n$, one may compute the mass ratios $\frac{m_{\tau}}{m_{\mu}}$ and $\frac{m_{\mu}}{m_e}$ from (8.8) in terms of the eigenvalues $\beta$ and $\gamma$ of $b$ and $c$ on the ground state $\ket{0}$.$^{(5)(6)}$

To obtain the neutrino masses one needs a conjugate Higgs doublet $ \begin{pmatrix} \rho_{\nu} \\ 0 \end{pmatrix}$.  Then

\begin{equation}
\left[  \begin{pmatrix} \bar{\nu}_L\bar{c}^3 & \bar{l}_L\bar{a}^3 \end{pmatrix}  \begin{pmatrix} \rho_{\nu}\\ 0 \end{pmatrix} \right] \left( \nu_Rc^3 \right) = \rho_{\nu} \left( \bar{c}^3 c^3 \right) \bar{\nu}_L \nu_R
\end{equation}

\noindent and
\begin{equation}
\bar{L}_{\nu}\varphi_{\nu}R_{\nu} + \bar{R}_{\nu}\bar{\varphi}_{\nu}L_{\nu} = m_{\nu}\bar{\nu}\nu
\end{equation}

\noindent where
\begin{equation}
m_{\nu}=\rho_{\nu}\bra{n}\bar{c}^3c^3\ket{n}
\end{equation}

The same discussion may be repeated for the up and down members of the quark doublet, and summarized by replacing the mass term of the standard model by$^{(5)(6)}$
\begin{equation}
\sum_i \sum_n \bra{n} \bar{L}(i)\varphi(i) R(i) + \bar{R}(i) \bar{\varphi}(i)L(i) \ket{n}
\end{equation}

\noindent where $n$ is summed over the three generations of each family and $i$ is summed over the four families: $i = (l, \nu, u, d)$.  The quark masses obtained in same way as (8.8) and (8.12) are

\begin{equation}
m_d = \rho_d \bra{n} \bar{b}^2 \bar{a} \cdot a b^2 \ket{n}
\end{equation}
\noindent and
\begin{equation}
m_u = \rho_u \bra{n} \bar{d}^2 \bar{c} \cdot c d^2 \ket{n}
\end{equation}

\section{The Fermion-Boson Interaction$^{(7)(8)}$}

In the standard model this interaction is expressed by
\begin{equation}
i(\bar{L}\nabla L + \bar{R}\nabla R)
\end{equation}

\noindent where $\nabla$ is the covariant derivative
\begin{equation}
\nabla = \slashed{\partial} + \slashed{\mathcal{W}}
\end{equation}

\noindent and $\mathcal{W}$ is the vector connection

\begin{equation}
\slashed{\mathcal{W}} = ig(\slashed{\mbox{W}}^+t_+ + \slashed{\mbox{W}}^-t_- + \slashed{\mbox{W}}^3 t_3) + i g_0 \slashed{\mbox{W}}^0 t_0
\end{equation}

\noindent We shall describe in detail only the non-Abelian contribution to $\mathcal{W}$.  (The Abelian $(g_o)$ term may be described in a simpler way.)

\noindent To go over to the SLq(2) model, replace (W$^+$, W$^-$, W$^3$) according to (7.3) by
\begin{equation}
(\mbox{W}^+ \mbox{D}^3_{-30}, \mbox{W}^-\mbox{D}^3_{30}, \mbox{W}^3\mbox{D}^3_{00})
\end{equation}

\noindent and replace
\begin{equation}
(\mbox{W}^+t_+, \mbox{W}^-t_-, \mbox{W}^3t_3)
\end{equation}

\noindent in (9.3) by
\begin{equation}
(\mbox{W}^+\mbox{D}^3_{-30}\cdot t_+, \mbox{W}^-\mbox{D}^3_{3 0}\cdot t_-, \mbox{W}^3\mbox{D}^3_{0 0}\cdot t_3)
\end{equation}
\noindent or by
\begin{equation}
(\mbox{W}^+\tau_+, \mbox{W}^-\tau_-, \mbox{W}^3\tau_3)
\end{equation}

\noindent where
\begin{eqnarray}
\tau_{\pm} &=& c_{\pm}t_{\pm}\mathcal{D}_{\pm} \\
\tau_3 &=& c_3 t_3 \mathcal{D}_3
\end{eqnarray}

\noindent Here the $\left( c_{\pm} , c_3 \right)$ are undetermined constants and

\begin{eqnarray}
\mathcal{D}_+ &=& c^3d^3 \hspace{0.1cm} ( \equiv \mbox{D}^3_{-30} \hspace{0.1cm} / \mbox{A}^3_{-30}) \\
\mathcal{D}_- &=& a^3b^3 \hspace{0.1cm} (\equiv \mbox{D}^3_{3 0} / \mbox{A}^3_{3 0} ) \\
\mathcal{D}_3 &=& f_3(b c) \hspace{0.1cm} (\equiv \mbox{D}^3_{0  0})
\end{eqnarray}

In defining $\mathcal{D}_+$ and $\mathcal{D}_-$ we may set
\begin{eqnarray}
d = \bar{a} \nonumber \\
c =\bar{b} \nonumber
\end{eqnarray}
i.e., we may identify the creation operators for the $d$ and $c$ preons with the creation operators for the antiparticles of the $a$ and $b$ preons, respectively, in agreement with Table. 4.5.

Then
\begin{displaymath}
\mathcal{D}_- = \overline{\mathcal{D}}_+
\end{displaymath}

\noindent The A$^3_{\pm3 0}$ numerical factors are absorbed in the $c_{\pm}$.  The $(c_{\pm},c_0)$ will be empirically determined in the next section 10.

The non-Abelian contribution to the covariant derivative in the knot model is now
\begin{equation}
\nabla = \slashed{\partial} + ig(\slashed{\mbox{W}}^+\tau_+ + \slashed{\mbox{W}}^-\tau_- + \slashed{\mbox{W}}^3\tau_3)
\end{equation}

\noindent and the non-Abelian part of the fermion-boson interaction in the knot Lagrangian is
\begin{equation}
\sum_i \sum_n \bra{n}\bar{L}(i) \nabla L(i)\ket{n}
\end{equation}

\noindent where L, R, and $\nabla$ are now all lying in the SLq(2) algebra, and where the sum over n is over the three generations, while the sum over i is over the two doublets. The only modification of $\nabla$ in going over to the knot model is the replacement of $\vec{t}$ by $\vec{\tau}$.

We next consider the detailed dependence of (9.14) on knot form factors.  For the lepton-neutrino doublet we have, dropping the Feynman slash,
\begin{equation}
\bar{L}\nabla L = \begin{pmatrix} \bar{\nu} & \bar{l} \end{pmatrix}_L(\partial + ig\mathcal{W}) \begin{pmatrix} \nu \\ l \end{pmatrix}_L
\end{equation}

\noindent The first term in (9.15) is
\begin{equation}
\tag{9.16a}
\begin{pmatrix} \bar{\nu} & \bar{l} \end{pmatrix}_L \partial \begin{pmatrix} \nu \\ l \end{pmatrix}_L = (\bar{c}^3 \bar{\nu}_L) \partial (c^3 \nu_L) + (\bar{a}^3 \bar{l}_L) \partial (a^3 l_L)
= (\bar{c}^3 c^3) \bar{\nu}_L \partial \nu_L + (\bar{a}^3 a^3)\bar{l}_L \partial l_L
\end{equation}

\noindent Here $ \begin{pmatrix} \nu \\ l \end{pmatrix}_L$ $\equiv$ $ \begin{pmatrix} c^3 \nu_L \\ a^3l_L \end{pmatrix}$ is the knot doublet and $\begin{pmatrix} \nu_L \\ l_L \end{pmatrix}$ is the doublet of the standard model.

Eqn. (9.16a) may be rewritten as

\begin{equation}
\tag{9.16b}
\bra{n} \begin{pmatrix} \bar{\nu} & \bar{l} \end{pmatrix}_L \partial  \begin{pmatrix} \nu \\ l \end{pmatrix}_L \ket{n} = \bar{\nu}_L\Delta_{\nu} \nu_L + \bar{l}_L \Delta_l l_L
\end{equation}
where
\begin{equation}
\tag{9.16c}
\Delta_{\nu}=\bra{n} \bar{c}^3 c^3 \ket{n} \partial \hskip1.0cm
\Delta_l = \bra{n} \bar{a}^3 a^3 \ket{n} \partial
\end{equation}
Then $\Delta_{\nu}$ and $\Delta_l$ are modified momentum operators rescaled with the same factors that rescale the neutrino and lepton rest masses found in the previous section.

\noindent The second term of (9.15) is by (9.8), (9.9), and (9.13)
\begin{eqnarray}
\addtocounter{equation}{1}
 \begin{pmatrix} \bar{\nu} & \bar{l} \end{pmatrix}_L \mathcal{W}  \begin{pmatrix} \nu \\ l \end{pmatrix}_L = \begin{bmatrix} \bar{c}^3\bar{\nu}_L & \bar{a}^3\bar{l}_L \end{bmatrix} \begin{bmatrix} c_3\mathcal{D}_3 \mbox{W}^3 & c_+\mathcal{D}_+\mbox{W}^+\\ c_-\mathcal{D}_-\mbox{W}^- & -c_3 \mathcal{D}_3 \mbox{W}^3 \end{bmatrix}  \begin{bmatrix} c^3\nu_L \\ a^3l_L \end{bmatrix} \\
= \begin{bmatrix} \bar{c}^3 \bar{\nu}_L & \bar{a}^3\bar{l}_L \end{bmatrix}  \begin{bmatrix} c_3\mathcal{D}_3\mbox{W}^3 \cdot c^3\nu_L + c_+\mathcal{D}_+\mbox{W}^+ \cdot a^3l_L \\ c_-\mathcal{D}_-\mbox{W}^- \cdot c^3\nu_L - c_3\mathcal{D}_3\mbox{W}^3 \cdot a^3 l_L \end{bmatrix} \nonumber \\
=c_3(\bar{c}^3 \mathcal{D}_3c^3)(\bar{\nu}_L \mbox{W}^3 \nu_L) + c_+(\bar{c}^3\mathcal{D}_+a^3)(\bar{\nu}_L\mbox{W}^+ l_L) \nonumber \\
+ \hspace{0.1cm} c_-(\bar{a}^3\mathcal{D}_-c^3)(\bar{l}_L\mbox{W}^-\nu_L) - c_3(\bar{a}^3\mathcal{D}_3a^3)(\bar{l}_L\mbox{W}^3 l_L)
\end{eqnarray}

There are four form factors stemming from the knot degrees of freedom, namely:

\begin{eqnarray}
\mbox{F}_{\bar{\nu}\nu} &=& c_3 \bra{n} \bar{c}^3\mathcal{D}_3c^3 \ket{n} = c_3 \bra{n} \bar{c}^3 f_3(b c) c^3 \ket{n} \\
\mbox{F}_{\bar{l}l} &=& c_3 \bra{n} \bar{a}^3 \mathcal{D}_3 a^3 \ket{n} = c_3 \bra{n} \bar{a}^3 f_3(b c) a^3 \ket{n} \\
\mbox{F}_{\bar{\nu}l} &=& c_+ \bra{n} \bar{c}^3 \mathcal{D}_+ a^3 \ket{n} = c_+ \bra{n} \bar{c}^3 (c^3d^3)a^3\ket{n} \\
\mbox{F}_{\bar{l}\nu} &=& c_- \bra{n} \bar{a}^3 \mathcal{D}_- c^3 \ket{n} = c_- \bra{n} \bar{a}^3 (a^3b^3) c^3 \ket{n} 
\end{eqnarray}

\noindent Here $f_3(bc) = \mbox{D}^3_{00}$ as in (9.12).  

Then the interaction is
\begin{equation}
\mbox{F}_{\bar{\nu} \nu}(\bar{\nu}_L\mbox{W}^3\nu_L) - \mbox{F}_{\bar{l}l}(\bar{\l}_L\mbox{W}^3 \l_L) + \mbox{F}_{\bar{\nu}l}(\bar{\nu}_L\mbox{W}^+l_L) + \mbox{F}_{\bar{l} \nu}(\bar{l}_L \mbox{W}^- \nu_L)
\end{equation}

\noindent All of these form factors are invariant under U$_a$(1)$\times$U$_b$(1) since $a$ and $d$, as well as $b$ and $c$, transform oppositely and each operator transforms oppositely to its adjoint.

For the up-down quark doublet we have
\begin{equation}
\bar{L} \nabla L =  \begin{pmatrix} \bar{u} & \bar{d} \end{pmatrix}_L (\partial + ig\mathcal{W}) \begin{pmatrix} u \\ d \end{pmatrix}_L
\end{equation}

\noindent where
\begin{equation}
 \begin{pmatrix} u \\ d \end{pmatrix}_L = \begin{pmatrix} cd^2 \cdot u_L \\ ab^2 \cdot d_L \end{pmatrix}
\end{equation}

\noindent Here again $ \begin{pmatrix} u \\ d \end{pmatrix}_L$ is the knot doublet while $ \begin{pmatrix} u_L \\ d_L \end{pmatrix}$ is the doublet in the standard model.

The first term of (9.24) is
\begin{equation}
\tag{9.26a}
\begin{array}{crll}  \begin{pmatrix} \bar{u} & \bar{d} \end{pmatrix}_L \partial \begin{pmatrix}u \\ d \end{pmatrix}_L & = & \overline{cd^2} \bar{u}_L \partial (cd^2)u_L + \overline{ab^2}\bar{d}_L\partial(ab^2)d_L \\ \hspace{2.0cm} & 
= & (\overline{cd^2} cd^2) \bar{u}_L \partial u_L + (\overline{ab^2} ab^2) \bar{d}_L \partial d_L \end{array}
\end{equation}

Eqn. (9.26a) may be rewritten as
\begin{equation}
\tag{9.26b}
\bra{n} \begin{pmatrix} \bar{u} & \bar{d} \end{pmatrix}_L \partial  \begin{pmatrix} u \\ d \end{pmatrix}_L \ket{n} = \bar{u}_L\Delta_u u_L + \bar{d}_L \Delta_d d_L
\end{equation}
where
\begin{equation}
\tag{9.26c}
\Delta_u=\bra{n} \overline{cd^2} cd^2 \ket{n} \partial  \hskip1.0cm \Delta_d = \bra{n} \overline{ab^2} ab^2 \ket{n} \partial
\end{equation}
Here $\Delta_u$ and $\Delta_d$ are modified momentum operators again rescaled with the same factors that rescale the rest masses of the $u$ and $d$ quarks in the previous section.

The second term is
\begin{eqnarray}
\addtocounter{equation}{1}
c_3[(\overline{cd^2})\mathcal{D}_3 (cd^2)] (\bar{u}_L \mbox{W}^3 u_L) + c_+[(\overline{cd^2})\mathcal{D}_+ (ab^2)](\bar{u}_L\mbox{W}^+d_L) \nonumber \\
+ \hspace{0.1cm} c_-[(\overline{ab^2})\mathcal{D}_-(cd^2)](\bar{d}_L\mbox{W}^-u_L) - c_3[(\overline{ab^2})\mathcal{D}_3(ab^2)](\bar{d}_L\mbox{W}^3d_L)
\end{eqnarray}

The interaction term in equation (9.24) is then the sum of four parts.
\begin{equation}
\mbox{F}_{\bar{u}u}(\bar{u}_L\mbox{W}^3 u_L) - \mbox{F}_{\bar{d} d} (\bar{d}_L\mbox{W}^3 d_L) + \mbox{F}_{\bar{u}d}(\bar{u}_L\mbox{W}^+ d_L) + \mbox{F}_{\bar{d}u} (\bar{d}_L\mbox{W}^- u_L)
\end{equation}

\noindent where the four form factors are
\begin{equation}
\mbox{F}_{\bar{u}u} = c_3 \bra{n} \overline{cd^2} f_3(bc) cd^2 \ket{n}
\end{equation}
\begin{equation}
\mbox{F}_{\bar{d}d} = c_3 \bra{n} \overline{ab^2}f_3(bc) ab^2 \ket{n}
\end{equation}
\begin{equation}
\mbox{F}_{\bar{u}d} = c_+ \bra{n} (\overline{cd^2}) \mathcal{D}_+ (ab^2) \ket{n} = c_+ \bra{n} \overline{cd^2} (c^3d^3) ab^2 \ket{n}
\end{equation}
\begin{equation}
\mbox{F}_{\bar{d}u} = c_- \bra{n} (\overline{ab^2}) \mathcal{D}_- (cd^2) \ket{n} = c_- \bra{n} \overline{ab^2} (a^3b^3) cd^2 \ket{n}
\end{equation}

\noindent All of these form factors are invariant under U$_a$(1) $\times$ U$_b$(1) since $a$ and $d$ transform oppositely as do $b$ and $c$.

After passing to SUq(2) all of the four form factors may be evaluated in terms of $q$ and $\beta$ where $\beta$ is the eigenvalue of $b$ on the ground state.

Since the R-fields are SU(2) singlets, they are invariant under SU(2) transformations and are not subject to SU(2) interactions.  They do transform according to hypercharge ($t_0$), or rotation charge.  These are U(1) gauge transformations, and $\bar{R} \nabla R$ is the sum of the following four parts:
\begin{equation}
\bra{n} \bar{c}^3 c^3 \ket{n}(\bar{\nu}_R (\partial + \mbox{W}^0) \nu_R)
\end{equation}
\begin{equation}
\bra{n}\bar{a}^3a^3\ket{n}(\bar{l}_R(\partial + \mbox{W}^0) l_R)
\end{equation}
\begin{equation}
\bra{n}\overline{cd^2} cd^2 \ket{n} (\bar{u}_R (\partial + \mbox{W}^0) u_R)
\end{equation}
\begin{equation}
\bra{n} \overline{ab^2} ab^2 \ket{n}(\bar{d}_R (\partial + \mbox{W}^0) d_R)
\end{equation}

\noindent All these terms are again invariant under U$_a$(1) $\times$ U$_b$(1) gauge transformations on the SLq(2) algebra.

\section{The Higgs Kinetic Energy Term$^{(7)(8)}$}

The weak neutral couplings are
\begin{equation}
i(g\mbox{W}_3\tau_3 +g_0\mbox{W}_0\tau_0) = i(\mathcal{A}\mbox{A} + \mathcal{Z}\mbox{Z})
\end{equation}

\noindent where the $t$ of the standard model has been replaced by $\tau$ as in $\left( 9.13\right)$ and
\begin{eqnarray}
\mbox{W}_0 &=& \mbox{A}\cos \theta - \mbox{Z} \sin \theta \\
\mbox{W}_3 &=& \mbox{A} \sin \theta + \mbox{Z}\cos \theta
\end{eqnarray}

\noindent Here $\theta$ is the Weinberg angle:
\begin{equation}
\tan \theta = \frac{g_0}{g}
\end{equation}

\noindent Then
\begin{eqnarray}
\mathcal{A}&=&g_0(\tau_3 + \tau_0)\cos\theta \\
\mathcal{Z}&=&g (\tau_3 -\tau_0 \tan^2\theta)\cos\theta
\end{eqnarray}

\noindent If $\ket{0}$ is a neutral state
\begin{equation}
\mathcal{A}\ket{0}=0
\end{equation}

\noindent By (10.5)
\begin{equation}
(\tau_3 + \tau_0)\ket{0}=0
\end{equation}

\noindent Then by (10.6)
\begin{equation}
\mathcal{Z}\ket{0}=\left( \frac{\tau_3}{\cos\theta} \right) \ket{0}
\end{equation}

\noindent Then by (10.1), (10.7), and (10.9)
\begin{equation}
\nabla( \varphi \ket{0})=\partial (\varphi \ket{0} + ig \left[ \mbox{W}_+ \tau_+ + \mbox{W}_- \tau_- + \frac{\mbox{Z}}{\cos\theta} \tau_3 \right] (\varphi \ket{0}) 
\end{equation}

\noindent and
\begin{equation}
\frac{1}{2}\bra{0}\mbox{Tr}\hspace{0.1cm} \overline{\nabla_{\mu}\varphi} \nabla^{\mu} \varphi \ket{0} = \frac{1}{2} \partial_{\mu} \rho \partial^{\mu} \rho + g^2 \rho^2 \left[ \mbox{I}_{++}\mbox{W}^{\mu}_+ \mbox{W}_{+ \mu} + \mbox{I}_{--}\mbox{W}^{\mu}_-\mbox{W}_{-\mu} + \frac{\mbox{I}_{33}}{\cos^2\theta}\mbox{Z}^{\mu}\mbox{Z}_{\mu} \right]
\end{equation}

\noindent where
\begin{displaymath}
\varphi = \left( \begin{array}{c} 0 \\ \rho \end{array} \right)
\end{displaymath}
and
\begin{equation}
\mbox{I}_{kk}=\frac{1}{2} \mbox{Tr} \hspace{0.1cm} \bra{0} \bar{\tau}_k \tau_k \ket{0} \hspace{2.0cm} k = +, -, 3
\end{equation}
and $\mbox{Tr}$ is taken over $\bar{t}_k t_k$.

\noindent To agree with the vector masses that are satisfactorily given by the standard model we have set $\bra{0}0\rangle = 1$ and shall also set
\begin{equation}
\mbox{I}_{kk}=\frac{1}{2} \mbox{Tr} \hspace{0.1cm} \bra{0} \bar{\tau}_k \tau_k \ket{0} = 1
\end{equation}

\noindent Since
\begin{equation}
\tau_k = c_k t_k \mathcal{D}_k \hskip1.0cm k = +,-,3
\end{equation}
the previously introduced and undetermined constants in (9.8) and (9.9) are now fixed by
\begin{equation}
\left|c_k\right|^{-2} = \frac{1}{2} \bra{0} \overline{\mathcal{D}}_k \mathcal{D}_k \ket{0}
\end{equation}
where the $\mathcal{D}_k$ are given by (9.10)-(9.12)

\noindent The $c_k$ are properly invariant and may be evaluated as functions of $q$ and $|\beta|^2$ in the same way that the form factors are evaluated in section 9.

We now replace the Higgs kinetic energy term of the standard model by
\begin{eqnarray}
\frac{1}{2} \mbox{Tr} \hspace{0.1cm} \bra{0} \overline{\nabla}_{\mu} \varphi \nabla^{\mu} \varphi \ket{0}
\end{eqnarray}
where $\nabla_{\mu} \varphi = \partial_{\mu} \varphi + ig[\mbox{W}_+\tau_+ + \mbox{W}_-\tau_- + \frac{\mbox{Z}}{\cos\theta}\tau_3]\varphi$ is the knot covariant derivative of a neutral scalar.

\section{Field Invariants $^{(7)(8)}$}

We replace the field invariant of the standard model by
\begin{equation}
\bra{0}\mbox{Tr}\hspace{0.1cm} \mathcal{W}_{\mu \lambda} \mathcal{W}^{\mu \lambda} \ket{0}
\end{equation}
where $\mathcal{W}_{\mu \lambda}$ are the field strengths of the knot model and where $\ket{0}$ is the ground state of the commuting $b$ and $c$ operators.

The covariant derivative is
\begin{equation}
\nabla_{\mu} = \partial_{\mu} + \mathcal{W}_{\mu}
\end{equation}
where $\mathcal{W}_{\mu}$ is the vector connection
\begin{equation}
\mathcal{W}_{\mu}=ig(\mbox{W}^+_{\mu}\tau_+ + \mbox{W}^-_{\mu}\tau_- + \mbox{W}^3_{\mu} \tau_3)
\end{equation}
where $\tau_{\pm}$ and $\tau_3$ are given by (9.8) - (9.12).

The field strengths are
\begin{eqnarray}
\mathcal{W}_{\mu \lambda} = \left[ \nabla_{\mu}, \nabla_{\lambda} \right]
= ig\left[ \partial_{\mu} \mbox{W}^p_{\lambda} - \partial_{\lambda} \mbox{W}^p_{\mu} \right] \tau_p - g^2 \mbox{W}_{\mu}^k \mbox{W}_{\lambda}^l \left[\tau_k, \tau_l \right]
\end{eqnarray}

\noindent and differ from the $\mbox{W}_{\mu}$ and $\mbox{W}_{\mu \lambda}$ of the standard model by the substitution of $\tau_k$ for $t_k$.

\noindent {\underline{The $\tau$-commutators lead to structure coefficients invariant under the gauge transformations}} {\underline{$U_a(1) \times U_b(1)$ that leave the SLq(2) algebra invariant and hence are functions of $bc$ only.}} The structure coefficients in (11.4) will therefore be functions of $\beta \gamma$, the value of $bc$ on the ground state in (11.1).

\section{Fermion and Preon Dynamics}

Interactions and masses of the fermions and preons are in principle determined by the Lagrangian described in the preceding section.  The fermions and preons, are described by the $\mbox{D}^{\frac{3}{2}}$ and $\mbox{D}^{\frac{1}{2}}$ representations, and interact by the $\mbox{D}^3$ vector bosons and by the $\mbox{D}^1$ vector bosons respectively.

 Since the number of preons equals the number of crossings, one may speculate that the crossings and preons are pointlike, that there is one preon at each crossing, and that the elementary fermions are composed of three preons bound by a trefoil of knot-electroweak and gluon fields.  If this is a realistic picture, there should be three bound states, corresponding to the three members of each family, with their observed masses; and assuming that the preon dynamics is entirely determined by the knotted action, the calculation of these bound states could be formulated as a well-defined mathematical problem.  On the other hand, to reach a physically credible picture, one needs some experimental guidance at relevant and presumably very high energies.  For example, one should expect the electroproduction of $a$ and $d$ particles according to
 \begin{center}
 $e^+ + e^- \rightarrow a + d + ...$
 \end{center}
 since they are charged $(\pm \frac{e}{3})$.
 The following decay modes are also kinematically possible:

\begin{quote}
Down quarks:
\begin{center}
$\mbox{D}^{\frac{3}{2}}_{\frac{3}{2}  -\frac{1}{2}} \rightarrow \mbox{D}^{\frac{1}{2}}_{\frac{1}{2} \frac{1}{2}} + \mbox{D}^1_{1  -1} \hspace{0.3cm} (ab^2 \rightarrow a + b^2)$
\end{center}
Up quarks:
\begin{center}
$\mbox{D}^{\frac{3}{2}}_{-\frac{3}{2} -\frac{1}{2}} \rightarrow \mbox{D}^{\frac{1}{2}}_{-\frac{1}{2}  \frac{1}{2}} + \mbox{D}^1_{-1  -1} \hspace{0.3cm} (cd^2 \rightarrow c + d^2)$
\end{center}
\end{quote}

These decays could limit to three the number of generations by permitting the quark to decay if given a critical dissociation energy.  In that case one would expect the formation of a preon-quark plasma at a sufficiently high temperature.
 
 Currently there is data at hadronic energies on electroweak reaction rates and on the masses of the three generations.  This data at present constrains and in principle is predicted by the knot model.  To discuss this data we now introduce some simplifications based on the same physical picture and on SUq(2), the unitary version of SLq(2).  Let us first consider the masses of the three generations of fermions.

\section{The Masses of the Fermions$^{(5)(6)}$}

The mass terms (8.8), (8.12), (8.14) and (8.15) of the knot Lagrangian contain the mass spectra of the four families that are listed in Table (13.1).$^{(5)(6)}$

\begin{center}
\underline{\bf{Table 13.1}}
\end{center}
\begin{center}
\begin{tabular}{c | c | c}
$i$ & $\mbox{D}^{\frac{3}{2}}_{mm'}(i)$ & $M(i, n)$ \\
\hline
$l$ & $a^3$ & $\rho(l) \bra{n} \bar{a}^3 a^3 \ket{n}$\\
$\nu$ & $c^3$ & $\rho(\nu) \bra{n} \bar{c}^3c^3 \ket{n}$\\
$d$ & $ab^2$ & $\rho(d) \bra{n} \bar{b}^2\bar{a} \cdot ab^2 \ket{n}$\\
$u$ & $cd^2$ & $\rho(u) \bra{n} \bar{d}^2\bar{c} \cdot cd^2 \ket{n}$\\
\end{tabular}
\end{center}
These masses are all of the form
\begin{displaymath}
\rho(m,m') \bra{n} \bar{\mbox{D}}^{\frac{3}{2}}_{mm'} \mbox{D}^{\frac{3}{2}}_{mm'} \ket{n}
\end{displaymath}
where $(m,m')$ labels the family and $n$ labels the generation.  The $\ket{n}$ are eigenstates of $\bar{\mbox{D}}^{\frac{3}{2}}_{mm'} \mbox{D}^{\frac{3}{2}}_{mm'}$.  Since the electric charge is $-\frac{e}{3}(m+m^{\prime})$, the pair $(m,m^{\prime})$ determines both the mass and the charge.

 In this table, as before, only the operator factor of the monomial $\mbox{D}^{\frac{3}{2}}_{mm'}$ is recorded.  The four prefactors $(\rho(l), \rho(\nu), \rho(d), \rho(u))$ represent the products of the numerical factors in $\mbox{D}^{\frac{3}{2}}_{mm'}$ with the Higgs factor.  The magnitude of $\rho$ sets the energy scale and differs for each family, possibly as required by the variation of the Yukawa coupling of the Higgs to each family.$^{(12)}$

The $M(i,n)$ in table 13.1 are invariant under U$_a$(1)$\times$U$_b$(1) transformations since the preon operators $(a, b, c, d)$ transform oppositely to their adjoints.

To numerically evaluate the expectation values of these operator products we go to the unitary version of SLq(2) by setting
\begin{eqnarray}
d&=&\bar{a} \\
c&=&-q_1\bar{b}
\end{eqnarray}
Then
\begin{equation}
\begin{array}{rcl} ab & = & qba \\ a\bar{b} & = & q\bar{b}a \end{array}
\hspace{1.0cm} 
\begin{array}{rcl} a\bar{a} + b\bar{b} & = & 1 \\ \bar{a}a + q_1^2\bar{b}b & = &1 \end{array}
\hspace{1.0cm} 
\begin{array}{rcl} b\bar{b} & = & \bar{b}b \\  {}\end{array}
\end{equation}

The identification of $d$ with $\bar{a}$ and $c$ with $\bar{b}$ is in agreement with the physical identification of the creation operators for the $d$ and $c$ preons with the creation operators, $\bar{a}$ and $\bar{b}$, for the antiparticles of the $a$ and $b$ preons respectively.  Then the operators $\bar{a}^na^n$ and $a^n\bar{a}^n$ are charge neutral and are expressible in terms of $b\bar{b}$ which is also charge neutral.

The reduction of $a^nd^n$ to a polynomial in $bc$ may be shown as follows:
\begin{eqnarray}
a^nd^n &=& a^{n-1} \cdot ad \cdot d^{n-1}\\
&=&a^{n-1}(1+qbc)d^{n-1} \nonumber\\ 
&=&a^{n-1}d^{n-1}(1+q^{2n-1}bc)
\end{eqnarray}

\noindent By iteration one finds
\begin{equation}
a^nd^n = \prod_{s=1}^{2n-1} \left(1+q^sbc \right)
\end{equation}
and in SUq(2)
\begin{equation}
a^n\bar{a}^n = \prod_{s=1}^{2n-1} \left(1-q^{s-1}b\bar{b} \right)
\end{equation}
and
\begin{equation}
\bar{a}^na^n = \prod_{s=1}^n \left(1 - q_1^{2s} b \bar{b} \right)
\end{equation}

We take the states $\ket{n}$ in Table 13.1 to be eigenstates of a mass operator expressed as a function of $b\bar{b}$. Then the expectation values for these states are functions of $\beta \bar{\beta}$, the eigenvalue of $b \bar{b}$ on the ground state. The $M \left(i,n \right)$ are then functions of $\left( q, \beta, n \right)$ and $\rho$, but the ratios of the masses in a single family depend only on $\left( q, \beta, n \right)$ and not on $\rho$.

The three generations, corresponding to the ground and two excited states, may be labelled by any three choices of $n$. The three expressions for the mass $M\left(i, n \right)$ correspond to the three choices of $n$ within a single family and are functions of the four parameters of the model $\left( q, \beta, n, \rho \right)$ according to
\begin{equation}
M \left(i,n\right) = \rho \left(i \right) F \left( q^2, |\beta|^2, n \right)
\end{equation}
where $F\left(q^2, |\beta|^2, n \right)$ is a polynomial in $|\beta|^2$ of the third degree, and a polynomial in $q^2$ of the degree determined by Table (13.1), eqns. (13.7) and (13.8) and the algebra (1.1).$^{(5)(6)}$

\noindent Depending on the assignment of $n$ to the three generations, one may determine $q$ and $\beta$ by eqn. (13.9) from the two ratios of the three observed masses. 



\section{Electroweak Reaction Rates$^{(6)(9)}$}

The matrix elements of the standard model acquire the following form factors in the corresponding knot model
\begin{equation}
\bra{n''} \overline{\mbox{D}}^{\frac{3}{2}}_{-3t''_3-3t''_0} \hspace{0.1cm} \mbox{D}^3_{-3t_3'-3t_0'} \hspace{0.1cm} \mbox{D}^{\frac{3}{2}}_{-3t_3-3t_0} \ket{n}
\end{equation}

\noindent where $n$ and $n''$ run over the three generations.

As an example consider
\begin{equation}
l^-+\mbox{W}^+ \rightarrow \nu_l
\end{equation}
with the following form factor
\begin{equation}
\bra{\nu_l} \overline{\mbox{D}}^{\frac{3}{2}}_{-\frac{3}{2} \frac{3}{2}} \hspace{0.1cm} \mbox{D}^3_{-3  0} \hspace{0.1cm} \mbox{D}^{\frac{3}{2}}_{\frac{3}{2} \frac{3}{2}} \ket{l}
\end{equation}
If this form factor is reduced in the SUq(2) algebra, it becomes a function of $q$ and $\beta$, where $\beta$ is the eigenvalue of $b$ on the ground state $\ket{0}$.  Comparison of (14.3) with experimental data on lepton-neutrino interactions like (14.2) indicates that 
\begin{equation}
(q, \beta) \cong (1, \frac{\sqrt{2}}{2})
\end{equation}
in approximate agreement with the universal Fermi interaction.$^{(6)}$

More demanding tests of the model are provided by the CKM and PMNS matrices that relate to (14.1), depending on whether $n$ and $n''$ label quarks or leptons and neutrinos.  In making these tests we introduce the further assumption that the flavor states are the ``coherent states", i.e., the eigenstates of the operators $\bar{a}$ and $a$, that are raising and lowering operators and thereby transmute particles of one generation into particles of the adjoining generation.$^{(9)}$

Starting from the mass states, one may obtain the flavor states as follows.  The orthonormal mass states $\ket{n}$ are defined to satisfy
\begin{eqnarray}
\ket{n} &=& \bar{a}^n \ket{0} \\
\bra{n} n' \rangle &=& \delta (n, n')
\end{eqnarray}
Then $\bar{a}$ is a raising operator:
\begin{equation}
\bar{a}\ket{n} = \lambda_n \ket{n+1}
\end{equation}
and
\begin{equation}
\bra{n}a = \lambda_n^* \bra{n+1}
\end{equation}

By (14.6) and (13.3)
\begin{equation}
|\lambda_n|^2 = 1 - q^{2n} | \beta |^2
\end{equation}
Similarly, $a$, working to the right on $\ket{n}$, is a lowering operator.  

Let $\ket{\alpha}$ be an eigenstate of $a$ with eigenvalue $\alpha$:
\begin{eqnarray}
a\ket{\alpha} &=& \alpha\ket{\alpha} \\
\bra{\alpha}\bar{a} &=& \bra{\alpha}\alpha^*
\end{eqnarray}
We now compute the matrix element $\bra{n} a \ket{\alpha}$ connecting mass and coherent states.

If $a$ operates to the right, one has by (14.10)
\begin{equation}
\bra{n} a \ket{\alpha} = \alpha \bra{n} \alpha \rangle
\end{equation}
and if it operates to the left, one has by (14.8)
\begin{equation}
\bra{n} a \ket{\alpha} = \lambda_n^*\bra{n+1} \alpha \rangle
\end{equation}
Then
\begin{equation}
\bra{n+1}\alpha \rangle = \frac{\alpha}{\lambda_n^*}\bra{n}\alpha \rangle
\end{equation}
By iteration,
\begin{equation}
\bra{n} \alpha \rangle = \frac{\alpha^n}{\prod_{0}^{n-1} \lambda_s^*} \bra{0} \alpha \rangle
\end{equation}
and $\bra{0}\alpha \rangle$ may be fixed by normalizing $\ket{\alpha}$.$^{(9)}$

Let the flavor states $\ket{i}$ be expressed as superpositions of the mass states $\ket{n}$:
\begin{equation}
\ket{i} = \sum \ket{n} \bra{n} i \rangle
\end{equation}
Then the matrix elements between flavor states are related to the matrix elements between mass states as follows
\begin{equation}
\bra{i} \mbox{M} \ket{i'} = \sum \bra{i} n \rangle \bra{n} \mbox{M} \ket{n'} \bra{n'} i' \rangle
\end{equation}
The mass states are orthonormal but the flavor (coherent) states are not orthogonal and their normalizations are also left free to be fixed by the data.

Let
\begin{equation}
\mbox{N}_i = \bra{i} i \rangle
\end{equation}
We are now interested in the generalization of (14.3) to the cases for which
\begin{equation}
\mbox{M} = \overline{\mbox{D}}^{\frac{3}{2}}_{m''p''} \hspace{0.1cm} \mbox{D}^3_{m'p'} \hspace{0.1cm} \mbox{D}^{\frac{3}{2}}_{mp}\hskip0.1cm\mbox{,}
\end{equation}
when taken between flavor states, describes the weak vector interactions of all the elementary fermions.  In particular
\begin{eqnarray}
\bra{u(i)} \mbox{W}^+ \ket{d(i')} &=& \sum_{nn'} \bra{u(i)} u(n) \rangle \bra{u(n)} \mbox{W}^+ \ket{d(n')} \bra{d(n')} d(i') \rangle \\
\bra{d(i)} \mbox{W}^- \ket{u(i')} &=& \sum_{nn'} \bra{d(i)} d(n) \rangle \bra{d(n)} \mbox{W}^- \ket{u(n')} \bra{u(n')} u(i') \rangle
\end{eqnarray}
holding for the flavor states of the up and down quarks.

\noindent With the same model for the PMNS matrix the form factor is
\begin{equation}
\bra{i}\bar{\mbox{D}}^{\frac{3}{2}}_{-\frac{3}{2}  \frac{3}{2}} \hspace{0.1cm} \mbox{D}^3_{0  0} \hspace{0.1cm} \mbox{D}^{\frac{3}{2}}_{-\frac{3}{2}  \frac{3}{2}} \ket{i'}
\end{equation}
where $i=0, 1, 2$ label the three generations of neutrino flavor states.

Because of the U$_m$(1) $\times$ U$_{m'}$(1) symmetry, the matrix element M in (14.19) is neutral, i.e., $n_a-n_d = n_b-n_c=0$.  It is therefore a function of $b$ and $c$ only and has no off-diagonal elements:

\begin{equation}
\bra{n} \mbox{M} \ket{n'} = \mbox{M}_n \delta(n,n')
\end{equation}
and by (14.17)
\begin{equation}
\bra{i} \mbox{M} \ket{i'} =\sum_n \bra{i} n \rangle \mbox{M}_n \bra{n} i' \rangle
\end{equation}

\noindent We propose that the quantities $\left| \bra{n} i \rangle \right|$ are the matrix elements of the Cabbibo-Kobayashi-Maskawa (CKM) matrix and may be expressed as a function of

\begin{quote}
(a) the eigenvalues of $a$: $\alpha$
\end{quote}
\begin{quote}
(b) the norms of the eigenstates of $a$: $N_i$
\end{quote}
\begin{quote}
(c) the matrix elements of $a$ between neighboring mass states: $\lambda_n = \bra{n} a \ket{n+1}$
\end{quote}
The matrix elements $\lambda_n$ are in turn functions of the two parameters $q$ and $\beta$ of the  model: $ \left| \lambda_n \right|^2 = 1 - q^{2n} \left|\beta \right|^2$

The elements $\bra{n} i \rangle$ are shown in Table 14.1$^{(9)}$:

\begin{center}
\underline{\bf{Table 14.1}}
\end{center}
\begin{center}
Elements of the $\bra{n} i \rangle$ Matrix
\end{center}
\begin{center}
\begin{tabular}{| c | c | c | c |}
\hline
n $\backslash$ i & 0 & 1 & 2\\
\hline
0 & $N_0^{\frac{1}{2}}g \left( \left| \frac{\alpha_0}{\lambda_0} \right| \right)$ & $ N_1^{\frac{1}{2}}g\left( \left| \frac{\alpha_1}{\lambda_0} \right| \right)$ & $N_2^{\frac{1}{2}}g \left( \left| \frac{\alpha_2}{\lambda_0} \right| \right)$ \\
\hline
1 & $N_0^{\frac{1}{2}} \frac{\alpha_0}{\lambda_0} g \left( \left| \frac{\alpha_0}{\lambda_0} \right| \right)$& $N_1^{\frac{1}{2}}\frac{\alpha_1}{\lambda_0} g \left( \left| \frac{\alpha_1}{\lambda_0} \right| \right)$ & $N_2^{\frac{1}{2}} \frac{\alpha_2}{\lambda_0}g \left( \left| \frac{\alpha_2}{\lambda_0} \right| \right)$\\
\hline
2 & $N_0^{\frac{1}{2}} \left( \frac{\alpha_0}{\lambda_0} \right)^2 \left( \frac{\alpha_0}{\lambda_1} \right)_0 g \left( \left| \frac{\alpha_0}{\lambda_0} \right| \right)$ & $N_1^{\frac{1}{2}} \left( \frac{\alpha_1}{\lambda_0} \right)^2 \left( \frac{\lambda_0}{\lambda_1} \right)_1g \left( \left| \frac{\alpha_1}{\lambda_0}\right| \right)$ & $N_2^{\frac{1}{2}} \left(\frac{\alpha_2}{\lambda_0} \right)^2 \left( \frac{\lambda_0}{\lambda_1} \right)_2 g \left( \left| \frac{\alpha_2}{\lambda_0} \right| \right)$\\
\hline
\end{tabular}
\end{center}

In this table $g(x_i) = \left[ 1 + x_i^2 + \left| \frac{\lambda_0}{\lambda_1} \right|^2 x_i^4 \right]^{-\frac{1}{2}}$ with $x_i = \frac{\alpha_i}{\lambda_0}$.

The absolute values $\left| \bra{n} i \rangle \right|$ are equated to the elements of the CKM matrix in Table 14.2:

\begin{center}
\underline{\bf{Table 14.2}}
\end{center}
\begin{center}
The CKM Matrix
\end{center}
\begin{center}
\begin{tabular}{| c | c | c | c |}
\hline
$n \backslash i$ & 0 & 1 & 2\\
\hline
0 & 0.97428 & 0.2253 & 0.00347\\
\hline
1 & 0.2252 & 0.97345 & 0.0410\\
\hline
2 & 0.00862 & 00403 & 0.999\\
\hline
\end{tabular}
\end{center}

We require a complete match between the $\left| \bra{n} i \rangle \right|$ taken from Table 14.1 with the entries in Table 14.2.  We then find $N_0 \cong N_1 \cong 1$ but $N_2 = 0.7$.  We also find a continuum of solutions for $q$ in the neighborhood of $q=1$.  These results are compatible with the conclusion that $q\cong1$ for the lepton-neutrino interactions.  

\newpage

\section{The Physical Interpretation of $``q"^{(10)}$}

In the present context $q$ is a parameter that measures the deformation of the standard model caused by the $``$knotting$"$ of the elementary fermion.  The empirical value of $q$ obtained from electroweak reaction rates is in the neighborhood of unity.  In particular, if the knot identification of the flavor states is accepted, then the observed CKM matrix indicates that the parameter, $q$, may be very close to unity.  On the other hand, if there is any SLq(2) substructure at all, the possibility that $q$ is precisely unity is excluded.

The primary substructure of quantum fields is determined by the Heisenberg algebra holding for the conjugate fields and realized by field quanta.  Here there is another substructure determined by the SLq(2) algebra and implemented by preons.

The Heisenberg and SLq(2) algebras may be related by the following quadratic form
\begin{equation}
\mbox{K} = \mbox{A}^t \varepsilon_q \mbox{A}
\end{equation}
where
\begin{equation}
\varepsilon_q = \begin{pmatrix} 0 & q^{-\frac{1}{2}} \\ -q^{\frac{1}{2}} & 0 \end{pmatrix} \hspace{2.0cm} \varepsilon_q^2 = -1
\end{equation}
This form is invariant under SLq(2) transformations of $\mbox{A}$.

\noindent Choosing
\begin{equation}
\mbox{A} = \begin{pmatrix} \mbox{D}_x \\ x \end{pmatrix} \hspace{1.7cm} \mbox{and} \hspace{1.7cm} \mbox{K} = q^{-\frac{1}{2}}
\end{equation}
one has by (15.1)$-$(15.3) the following SLq(2) invariant relation
\begin{equation}
\mbox{D}_x x - qx\mbox{D}_x = 1
\end{equation}

Equation (15.4) is identically satisfied if $\mbox{D}_x$ is chosen as the $q$-difference operator, namely
\begin{equation}
\mbox{D}_x \Psi(x) = \frac{\Psi(qx)-\Psi(x)}{qx-x}
\end{equation}
If we introduce
\begin{equation}
\mbox{P}_x = \frac{\hbar}{i} \mbox{D}_x
\end{equation}
then we have the SLq(2) invariant relation
\begin{equation}
\left( \mbox{P}_xx-qx\mbox{P}_x \right) \Psi(x) = \frac{\hbar}{i} \Psi(x)
\end{equation}
If $q \rightarrow 1$ then (15.7) becomes the Heisenberg commutator applied to a quantum state.  Otherwise D$_x$ resembles, by (15.5), the differentiation operator on a lattice space and $q$ may play the role of a dimensionless regulator.

In view of the physical evidence suggestive of substructure, which has been described here, as well as the natural appearance of the non-standard $q$-derivative, it may be possible to utilize SLq(2) to describe a finer level of structure than is currently considered.

We have ignored the gravitational field in this paper since it is not immediately relevant.  As we have, however, discussed the knot symmetries of the fundamental particles, we have thereby also discussed the knot symmetries of these sources of the gravitational field.  Since one expects that the symmetries of its source would in some measure be inherited by the gravitational field itself, it is interesting that knot states have emerged in a natural way in attempts to quantize general relativity.$^{(11)}$

\section{Acknowledgements}

I thank J. Smit, A.C. Cadavid, and J. Sonnenschein for helpful discussion.

\newpage
$\Large{\bf{References}}$
\begin{enumerate}
\item Thomson, W. H., Trans. R. Soc. Edinb. $\bf{25}$, 217-220 (1969)
\item Faddeev, L. and Niemi, Antti, J., Nature $\bf{387}$, May 1 (1997)
\item Finkelstein, R.J., Knots and Preons,  Int. J. Mod. Phys. A$\underline{\bf{24}}$, 2307 (2009)
\item Finkelstein, R.J., Colored Preons,  arXiv:0901.1687
\item Finkelstein, R.J., A Knot Model Suggested by the Standard Electroweak Theory, Int. J. Mod. Phys. A $\underline{\bf{20}}$ 6487 (2005)
\item Cadavid, A.C. and Finkelstein, R.J., Masses and Interactions of q-Fermionic Knots,  Int. J. Mod. Phys. A $\underline{\bf{21}}$ 4269 (2006)
\item Finkelstein, R.J., Field Theory of Quantum Knots,  [hep.th] 0701124
\item Finkelstein, R.J., Solitonic Models Based on Quantum Groups and the Standard Model, arXiv:1011.2545 v1 [hep.th]
\item Finkelstein, R.J., Flavor States of the Knot Model, arXiv:1011.0764 v3 [hep.th]
\item Finkelstein, R.J., On the Deformation Parameter in SLq(2) Models of the Elementary Particles, arXiv:1108.0438 v1 [hep.th]
\item Horowitz, G.T., Strings and Symmetries, World Scientific (1991)
\item J. Sonnenschein, unpublished
\item H. Harari, Physics Letters B $\underline{\bf{86}}$, 83-86, M. Shupe, Physics Letters B $\underline{\bf{86}}$, 87-92
\end{enumerate}

\end{document}